\documentclass[%
 reprint,
superscriptaddress,
 amsmath,amssymb,
 aps,
prb,
]{revtex4-1}

\usepackage{graphicx}% Include figure files
\usepackage{dcolumn}% Align table columns on decimal point
\usepackage{bm}% bold math
\usepackage{soul}
\usepackage{hyperref}% add hypertext capabilities
\hypersetup{colorlinks=true,citecolor={blue},linkcolor={blue},urlcolor={blue}}
\usepackage{color}
\usepackage[dvipsnames]{xcolor}
\usepackage{cancel}
\usepackage{ulem}

\newcommand{\diag}{\mathop{\mathrm{diag}}\nolimits}
  \newcommand{\J}{\widehat{\mathbb{J}}}
  \newcommand{\jhat}[1]{\widehat{\mathcal{J}}_{#1}}
  \renewcommand{\O}{\widehat{\mathbb{O}}}
  \renewcommand{\vec}[1]{\mathbf{#1}}
  \newcommand{\sigmai}[1]{\widehat{\sigma}_{#1}}
  \newcommand{\sigmax}{\sigmai{x}}
  
  \newcommand{\sigmaz}{\sigmai{z}}
  \newcommand{\sigmao}{\sigmai{0}}
  \newcommand{\Shati}[1]{\widehat{S}_{#1}}
  \newcommand{\alphaave}{\overline{\alpha}}
  \newcommand{\e}{\mathrm{e}}
  \newcommand{\eo}{\vec{\mathtt{e}}_0}
  \newcommand{\eplus}{\e_+}
  \newcommand{\eminus}{\e_-}
  \newcommand{\rhohalf}{\frac{\sqrt{\rho}}{2}}
  \newcommand{\Htot}{\widehat{H}}
  \newcommand{\Hzero}{\widehat{H}_0}
  \newcommand{\Hbulk}{H_{\text{bulk}}}
  \newcommand{\Ezero}{\widehat{\mathbb{E}}_0}
  \newcommand{\Splus}{\widehat{\mathbb{S}}_+}
  \newcommand{\Sminus}{\widehat{\mathbb{S}}_-}

  \newcommand{\abs}[1]{\left| #1 \right|}
  \newcommand{\br}[1]{\left( #1 \right)}
  
  \newcommand{\ea}{\vec{e}_A}
  \newcommand{\uvplus}{e_A^+ v_A^+ + e_A^{+*} u_A^+}
  \newcommand{\uvminus}{e_A^- v_A^- + e_A^{-*} u_A^-}
  \newcommand{\uvplusc}{e_A^+ u_A^{+*} + e_A^{+*} v_A^{+*}}
  \newcommand{\uvminusc}{e_A^- u_A^{-*} + e_A^{-*} v_A^{-*}}
  \newcommand{\alphaetaone}{\alpha_1 - \frac{i\eta}{2}}
  \newcommand{\alphaetatwo}{\alpha_2 - \frac{i\eta}{2}}
  \newcommand{\alphaetaoneplus}{\alpha_1 + \frac{i\eta}{2}}

\newcommand{\ns}{\mathcal{N}}

\begin{document}

\preprint{APS/123-QED}

\title{Spontaneous topological transitions in a honeycomb lattice of exciton-polariton condensates due to spin bifurcations}

\author{H. Sigurdsson}
\affiliation{School of Physics and Astronomy, University of Southampton, SO17 1BJ, Southampton, United Kingdom}
\affiliation{Skolkovo Institute of Science and Technology, Skolkovo Innovation Center, Building 3, Moscow 143026, Russian Federation}

\author{Y. S. Krivosenko}
\affiliation{ITMO University, St. Petersburg 197101, Russia}

\author{I. V. Iorsh}
\affiliation{ITMO University, St. Petersburg 197101, Russia}

\author{I. A. Shelykh}
\affiliation{Science Institute, University of Iceland, Dunhagi 3, IS-107, Reykjavik, Iceland}
\affiliation{ITMO University, St. Petersburg 197101, Russia}

\author{A. V. Nalitov}
\affiliation{Science Institute, University of Iceland, Dunhagi 3, IS-107, Reykjavik, Iceland}
\affiliation{ITMO University, St. Petersburg 197101, Russia}
\affiliation{Faculty of Science and Engineering, University of Wolverhampton, Wulfruna Street, WV1 1LY, Wolverhampton, United Kingdom}

\date{\today}

\begin{abstract}
We theoretically study the spontaneous formation of the quantum anomalous Hall effect in a graphene system of spin-bifurcated exciton-polariton condensates under nonresonant pumping. We demonstrate that, depending on the parameters of the structure, such as intensity of the pump and coupling strength between condensates, the system shows rich variety of macroscopic magnetic ordering, including analogs of ferromagnetic, antiferromagnetic, and resonant valence bond phases. Transitions between these magnetic polarized phases are associated with dramatic reshaping of the spectrum of the system connected with spontaneous appearance of topological order.
\end{abstract}

\maketitle

%\tableofcontents
\textit{Introduction}.
Recent decades have witnessed a shift in attention from both the condensed matter and optical communities in investigation of the properties of bulk materials to instead the properties of their interfaces. It is now well known that there exists a particular class of materials with inverted structure of the bands, which possess protected states propagating on the system surface, referred to as topological insulators \cite{Hasan2010, Qi2011, Bansil2016}. The energy of these edge states lies within the bandgap, and thus they are protected with respect to scattering into the bulk. Depending on the dimensionality of a system, one should distinguish between three-dimensional (3D) topological insulators where topological states appear on the two-dimensional (2D) surface boundary of the bulk \cite{Kane2005a,Fu2007,Moore2007,Roy2009,Xia2009,Zhang2009,Chen2009}, and 2D topological insulators where chiral 1D channels form on the system boundary \cite{Bernevig2006,Konig2007}.

It has been recently shown that optical analogs of topologically nontrivial phases, in 2D systems, may arise in purely photonic structures~\cite{Haldane2008, Wang_Nature2009, Khanikaev2013} or when driven into the strong light-matter coupling regime, where hybrid quasiparticles known as cavity exciton-polaritons (from here on {\it polaritons}) are formed~\cite{Karzig-PRX-2015, Bardyn-PRB-2015, Nalitov-Z, KagomePolariton, Kozin2018, Klembt_Nature2018}. Polaritons combine the advantages of photons, such as extremely low effective mass and long coherence length, with those of excitons, namely the possibility of control by external electric and magnetic fields together with strong nonlinear response stemming from interparticle interactions. The spin structure of the exciton (or rather, the polariton) is then directly related to the circular polarization degree of the cavity photonic mode. Such a union then leads to a rich interplay between nonlinear and topological properties~\cite{Bleu2016, Bardyn_2016PRB, Gulevich2017, Sigurdsson2017, Kartashov2017, Bleu_PRB2017, Mandal_PRB2019} with optical lasing in topologically protected edge modes~\cite{StJean_NatPho2017, Klembt_Nature2018}.

The vast majority of current proposals on 2D polariton topological insulators are based on $\mathbb{Z}$ (or Chern) insulators with the following requirements: Polaritons should be placed into a 2D lattice of a particular symmetry allowing the appearance of Dirac points in the Brillouin zone where the bands touch each other. Examples are honeycomb \cite{Nalitov-Z, Klembt_Nature2018, Kartashov_PRL2019} and Kagome \cite{KagomePolariton, Gulevich2017, Sigurdsson2017} lattices which can be obtained either by controllable etching of a planar microcavity or by using spatial light modulator to control the profile of the external optical pump. The band inversion and opening of the topological gap is then achieved by cumulative action of the TE-TM splitting of the photonic mode and Zeeman splitting of the excitonic mode induced by the application of an external magnetic field~\cite{Nalitov-Z}. However, in conventional semiconductor materials excitonic $g$-factors are extremely small, and one needs magnetic fields of tens of Tesla to open the topological bandgap of at least several meV. The situation can be potentially improved by using diluted magnetic microcavities \cite{Brunetti2006,Krol2018}. However, the technology of producing a high quality patterned semimagnetic cavity is still only in its initial stages.

In the present paper, we develop an alternative approach for the realization of the quantum anomalous Hall effect in a 2D polariton $\mathbb{Z}$-topological insulator without application of any external magnetic fields. Our idea is based on the concept of the spontaneous spin bifurcation in a system of interacting polariton condensates forming a net magnetic polarization, first proposed in Ref.~[\onlinecite{Ohadi2015}] and developed further in Refs.~[\onlinecite{Ohadi2016, Dreismann2016, Ohadi2017, Sigurdsson2017a}]. Here we consider a honeycomb polariton condensate lattice (polariton graphene) under nonresonant pumping. We demonstrate that, depending on the pump intensity and coupling strength between the nodes (condensates) of the lattice, the spin bifurcation mechanism can result in spontaneous formation of distinct spin-ordered lattice phases analogous to ferromagnetic (FM), antiferromagnetic (AFM), and resonance valence bond states. Transition between different phases is associated with cardinal reshaping of the spectrum of the system excitations and spontaneous appearance/disappearance of topological order.

\textit{Spin bifurcations and phase transitions in polariton graphene}.
A lattice of driven-dissipative connected polariton condensates is conventionally modeled, in the tight-binding picture, with a set of generalized Gross-Pitaevskii equations for the spinor order parameters $\Psi_n = (\psi_{n+}, \psi_{n-})^\text{T}$, corresponding to spin-up and spin-down polaritons at the $n$-th site,
\begin{align} \notag
        & i \frac{d\Psi_n}{dt}  =
        \left[ - {i \over 2}g(S_n)
        -{\epsilon+ i \gamma \over 2} \hat{\sigma}_x  + {1\over2} \left( \bar{\alpha} S_n + \alpha S_n^z \hat{\sigma}_z \right) \right] \Psi_n
        \\  \label{eq.GP}
        &  -\frac{1}{2} \sum_{\langle nm \rangle} {\left[ J +  \delta J \left( \cos (2\varphi_m) \hat{\sigma}_x + \sin(2 \varphi_m) \hat{\sigma}_y \right) \right]  \Psi_m}.
\end{align}
Here the summation is taken over the nearest neighbors, $\varphi_m$ are the angles of links connecting the neighboring sites $n$ and $m$ of the honeycomb lattice.
We define the $n$-th node particle population $S_n$ and  $z$-component of the condensate pseudospin $S_n^z$ as:
\begin{equation} \label{eq.Sn}
S_n \equiv \frac{|\psi_{n+}|^2 + |\psi_{n-}|^2}{2}, \; S_{n}^z \equiv \frac{|\psi_{n+}|^2 - |\psi_{n-}|^2}{2}.
\end{equation}
We also define an effective decay rate $g(S_n) = \eta S_n +\Gamma - W$ with $\Gamma$ being the polariton decay rate, $W$ the replenishment rate of the condensate non-polarized incoherent pump, and $\eta$ is the gain-saturation nonlinearity.
The constants $\epsilon$ and $\gamma$ define the splitting of the $XY$-polarized states in both energy and decay, respectively, due to the inherent cavity birefringence, and $\bar{\alpha} = \alpha_1 + \alpha_2$ and $\alpha = \alpha_1 - \alpha_2$ are  spin-anisotropic interaction parameters. Finally, $J > \delta J$ are spin conserving and non-conserving (TE-TM splitting) tunneling rates of polaritons between nodes respectively.

\begin{figure}
\centering
\includegraphics[width=0.9\linewidth]{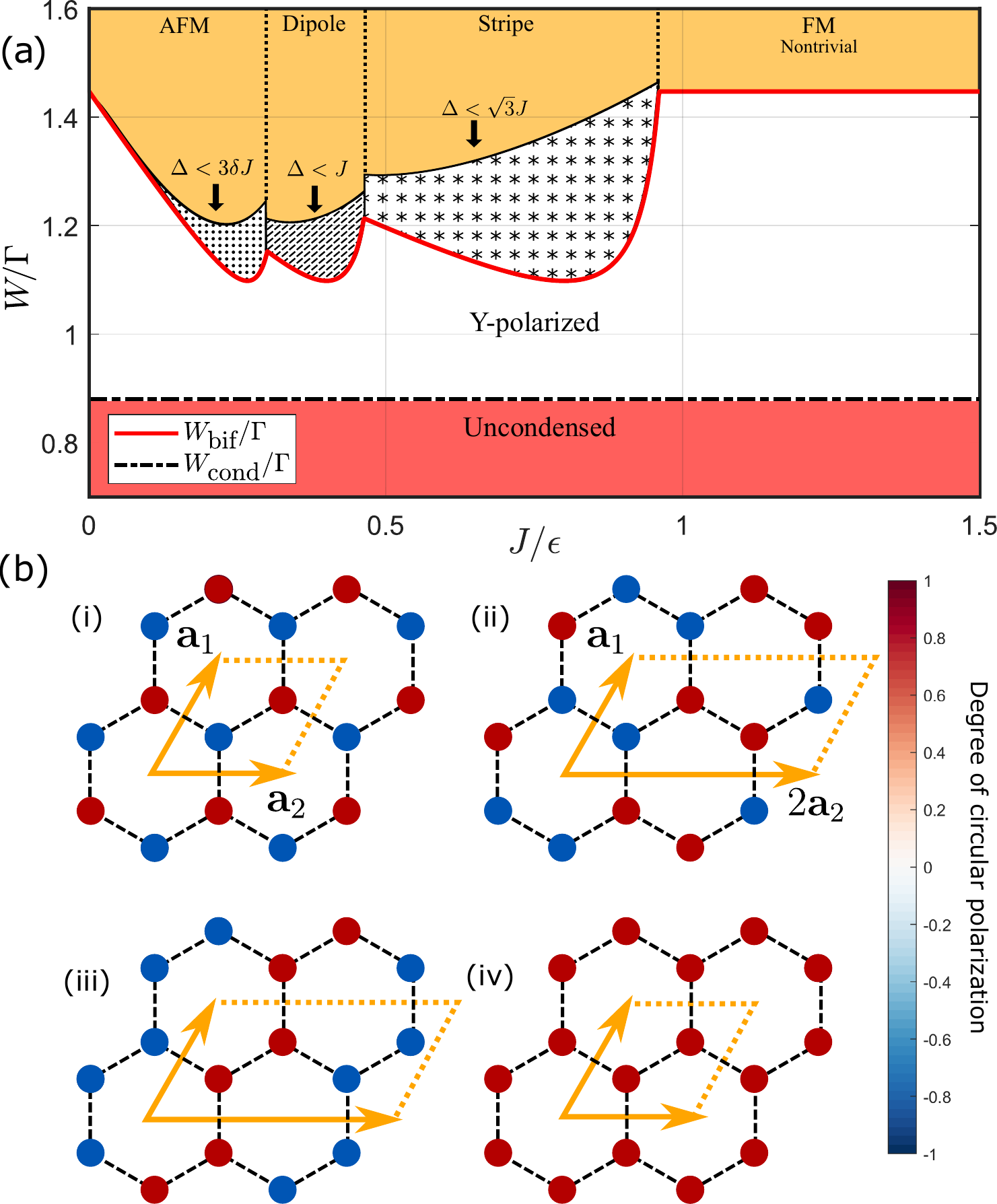}
\caption{(a) Phase map of the honeycomb polariton system. At low pump powers $W$ no condensation occurs (red area). When crossing the condensation threshold (dashed-dotted line) the system condenses and settles into $Y$-polarized state (white area). At higher powers the system bifurcates into an organized spin pattern (above red curve). Bandgap opening in each phase takes place in the yellow area whereas in white patterned areas the bandgap is closed. Only in the FM phase is the gap topologically nontrivial. (b) Cutout pieces of the four spin graphene patterns labeled as (i) AFM, (ii) dipole, (iii) stripe, and (iv) FM. Yellow arrows denote the unit cell vectors of each pattern. Parameters are chosen based on experiment~\cite{Ohadi2015}:  $\eta=0.005$  ps$^{-1}$;  $\Gamma=0.1$ ps$^{-1}$;  $\epsilon=0.06$ ps$^{-1}$;  $\gamma=0.2\epsilon$; $\alpha_1= 0.005$ ps$^{-1}$;  $\alpha_2=-0.1\alpha_1$.}
\label{fig1}
\end{figure}

The condensation threshold of the system is defined as the point where an eigenvalue of the linearized Eq.~\eqref{eq.GP} obtains a positive imaginary component due to increase of the laser power $W$, leading to the triggering of the stimulated bosonic scattering into the condensed state at $W_\text{cond} = \Gamma - \gamma$. Due to the splitting $\gamma$ in the lifetimes of the linear polarized states the condensate first forms an, in-phase, $Y$-polarized state, i.e., $\Psi_n = \Psi_{n+1} \propto (1,-1)^T$ (white area in Fig.~\ref{fig1}a). This $Y$-polarized state, however, becomes unstable at higher pumping powers and undergoes a bifurcation into a state with a high degree of circular polarization at individual nodes~\cite{Ohadi2015, Ohadi2017}.

We begin our consideration by presenting a class of stationary solutions which minimize the spin bifurcation threshold~\cite{Sigurdsson2017a},
\begin{equation} \label{ansatz}
    \Psi_n = \left\{
  \begin{array}{l l}
\Psi_{n+1}, & \quad \text{if} \ \ S_n^z = S_{n+1}^z, \\
- \hat{\sigma}_x \Psi_{n+1}, & \quad \text{if} \ \  S_n^z = -S_{n+1}^z. \\
\end{array} \right.
\end{equation}
The ansatz above describes in-phase FM bonds and anti-phase AFM bonds between nearest neighbors respectively. Plugging Eq.~\eqref{ansatz} into Eq.~\eqref{eq.GP}, and setting the condition that all nodes have the same number of co- and counterpolarized nearest neighbours (equivalence criteria), the coupled set of the equations of motion reduce to a single equation with a bifurcation threshold,
\begin{equation} \label{eq.bif}
W_\text{bif} = \Gamma - \gamma + \eta \frac{(\epsilon - n_{\uparrow \downarrow} J)^2 + \gamma^2}{\alpha (\epsilon - n_{\uparrow \downarrow} J) },
\end{equation}
\begin{figure*}[t!]
  \centering
  \includegraphics[width=\linewidth]{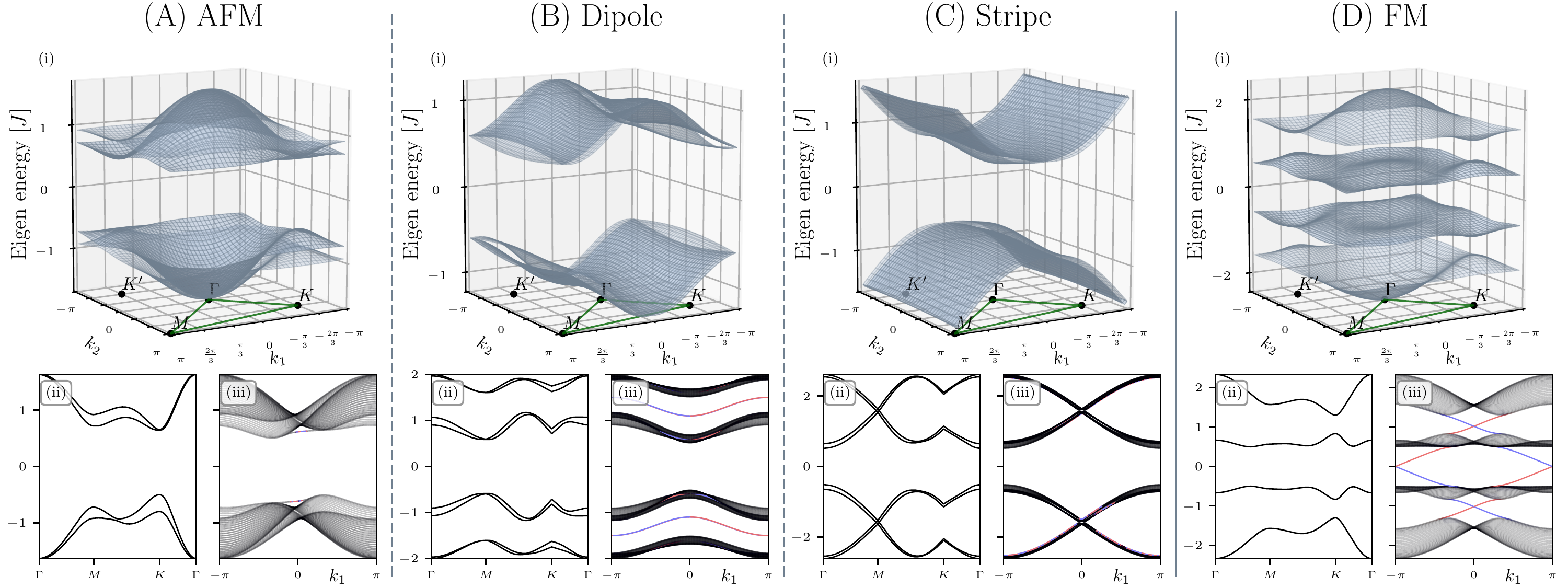}
  \caption{(i) The band structure of the whole Brillouin zone for the four spin phases in the honeycomb lattice of spin bifurcated polariton condensates. Here, $J = 1$ is taken as the unit of energy. (A, D): all the four bands are displayed, (B, C): only two bands above and two bands under the midgap are presented. Bottom panels show (ii) the eigenenergies along $\Gamma \to M \to K \to \Gamma$ pathway, and (iii) band structure of the ribbon with zigzag edges, the edge states are depicted by red and green dots. Parameters for the AFM: $\delta J = 0.1J$, $\Delta = 1.3J$. Dipole: $\delta J = 0.1J$, $\Delta = 2.2J$. Stripe: $\delta J = 0.1J$, $\Delta = 3J$. FM: $\delta J = 2J/3$, $\Delta = 5J/3$. The band structures of the bulk were calculated on a $200 {\times} 200$ mesh grid in $k$-space. The size of ribbons was $30$ (width) by $100$ (length) unit cells.}
  \label{fig:AFM_BS}
\end{figure*}
where $n_{\uparrow \downarrow}$ denotes the number of AFM neighbors. 

In Fig.~\ref{fig1}a we plot the minimum of Eq.~\eqref{eq.bif} as a function of coupling strength $J$ (red curve) neglecting TE-TM splitting. The cusps in the red curve indicate that the lowest bifurcation point is shared between two distinct spin phases. The four spin phases of interest, verified by numerical integration of Eq.~\eqref{eq.GP}, are shown in  Fig.~\ref{fig1}b(i-iv) and are labeled {\it AFM}, {\it Dipole}, {\it Stripe}, and {\it FM} phases respectively. 

We point out that the bifurcation threshold for AFM and FM phases is invariant of $\delta J$ whereas for dipole and stripe phases, strictly speaking, this is not the case as for them the ansatz given by the Eq.~\eqref{ansatz} should be modified. However, given that $\delta J /J \ll 1$ (which is usually the case in micropillar structures in standard semiconductor microcavities), it is reasonable to infer that $W_\text{bif}$ is only weakly affected by $\delta J$. Thus, the calculated red curve shown in the Fig.~\ref{fig1}a serves as a good indicator for the bifurcation threshold of these nontrivial states in the presence of TE-TM splitting. We have performed numerical calculations of Eq.~\eqref{eq.GP} that verify that this is indeed the case.

\textit{Band structure and topological states}.
In the following we discuss the excitation spectra of the stable spin bifurcated condensate configurations.
We employ an effective field model, treating the effect of spin-polarized lattice nodes by introduction of local $z$-directed (out of the cavity plane) magnetic fields. The idea is based on expansion of the Bogoliubov dispersion in the two spin components $\sqrt{E_\pm(E_\pm+ U_\pm)}\approx E_\pm+U_\pm/2$ with $U_\pm \approx \alpha_1 | \psi_{n \pm } |^2$ being the self interaction energy. It yields the effective magnetic field magnitude $\Delta = \alpha_1 |S_n^z|$ and the validity region of the approximation as $E_\pm\gg U_\pm$. The effective field model allows clear analytical investigation of the topological properties of the lowest band gap opening at $E \sim J$ as long as $J \gg \Delta$.
Furthermore, as the pseudospin strength is given by $|S_n^z| = S_n = (W-\Gamma)/\eta$ for the fully polarized condensates (see Eq.~(S17) in~[\onlinecite{Supplementary2019}]) one can adjust the strength of $\Delta$ such that effects due to birefringence are negligible. The effective field model then obeys $J \gg \Delta \gg \epsilon$. A more rigorous Bogoliubov treatment is addressed in Ref.~[\onlinecite{Supplementary2019}].

Firstly, to examine the band structures of the AFM (i) and FM (iv) configurations shown in Fig.~\ref{fig1}b, we scrutinize  the following $4{\times}4$ tight-binding Hamiltonian:
\newcommand{\Jhat}{\hat{\mathcal{J}}} % \hat should be finally changed to \widehat, as the latter looks better after an appropriate compilation
\newcommand{\Zero}{\mathbb{O}}
\renewcommand{\vec}[1]{\mathbf{#1}}
\newcommand{\Jtot}{\widehat{\mathbb{J}}}
\begin{equation}\label{eq:Hamiltonian4x4inK}
  H_{\vec{k}} = - \frac{1}{2}
  \begin{pmatrix}
    0 & \Jhat_{\vec{k}} \\
    \Jhat_{\vec{k}}^\dagger & 0
  \end{pmatrix}
  + \frac{1}{2}
  \begin{pmatrix}
    \mu_1 \Delta \, \hat{\sigma}_z & 0 \\
    0 & \mu_2 \Delta \, \hat{\sigma}_z
  \end{pmatrix}.
\end{equation}
The above Hamiltonian is written in the basis of the bispinor $| \text{A}+, \text{A}-, \text{B}+, \text{B}-\rangle$ inner-cell states, where A and B indicate the graphene sublattices, and ``$+(-)$'' specifies right (left) circular polarization (i.e., the spin of the polaritons). The total Hamiltonian in momentum space is then  written $\hat{H} = \sum_\vec{k} | \vec{k} \rangle \langle \vec{k} | \otimes H_{\vec{k}}$. The first term in Eq.~\eqref{eq:Hamiltonian4x4inK} corresponds to the polaritonic graphene with TE-TM splitting.
The $2{\times}2$ operator $\Jhat_{\vec{k}}$, dependent on the quasi-wavevector $\vec{k} = (k_1, k_1)$, is written,
\begin{equation}{\label{eq:J_operator}}
  \Jhat_{\vec{k}} = \jhat{1} + \jhat{2} \, \e^{-i k_1} + \jhat{3} \, \e^{-i k_2},
\end{equation}
with
\begin{equation}
  \jhat{m} =
  \begin{pmatrix}
    J & \delta J \, \e^{-2i \varphi_m} \\
    \delta J \, \e^{2i \varphi_m} & J
  \end{pmatrix}, \quad m = 1, 2, 3.
\end{equation}
and $J$ and $\delta J$ as in Eq.~\eqref{eq.GP}. Both $k_1$ and $k_2$ can be chosen to vary from $-\pi$ to $\pi$ and to cover the whole Brillouin zone.
In the second term of Eq.~\eqref{eq:Hamiltonian4x4inK}, the on-diagonal blocks $\mu_{1(2)} \Delta \, \sigma_z$ serve to account for the excitations in the magnetic patterns depicted in Fig.~\ref{fig1}b.
The magnitude of the Zeeman splitting induced by the polarized condensate reads $\Delta = \alpha_1 S$ where we have omitted the index $n$ in $S_n$ since the condensate density is taken equal at each lattice site. Here $\hat{\sigma}_z$ is the $z$-Pauli matrix, and the coefficients $\mu_1$ and $\mu_2$ define FM ($\mu_1 = \mu_2 = 1$) and AFM ($\mu_1 = -\mu_2 = 1$) phases. The translation basis vectors ($\mathbf{a}_{1,2}$) in real space are chosen to be conventional for the graphene lattice.
\begin{figure}[t!]
  \centering
  \includegraphics[width=\columnwidth]{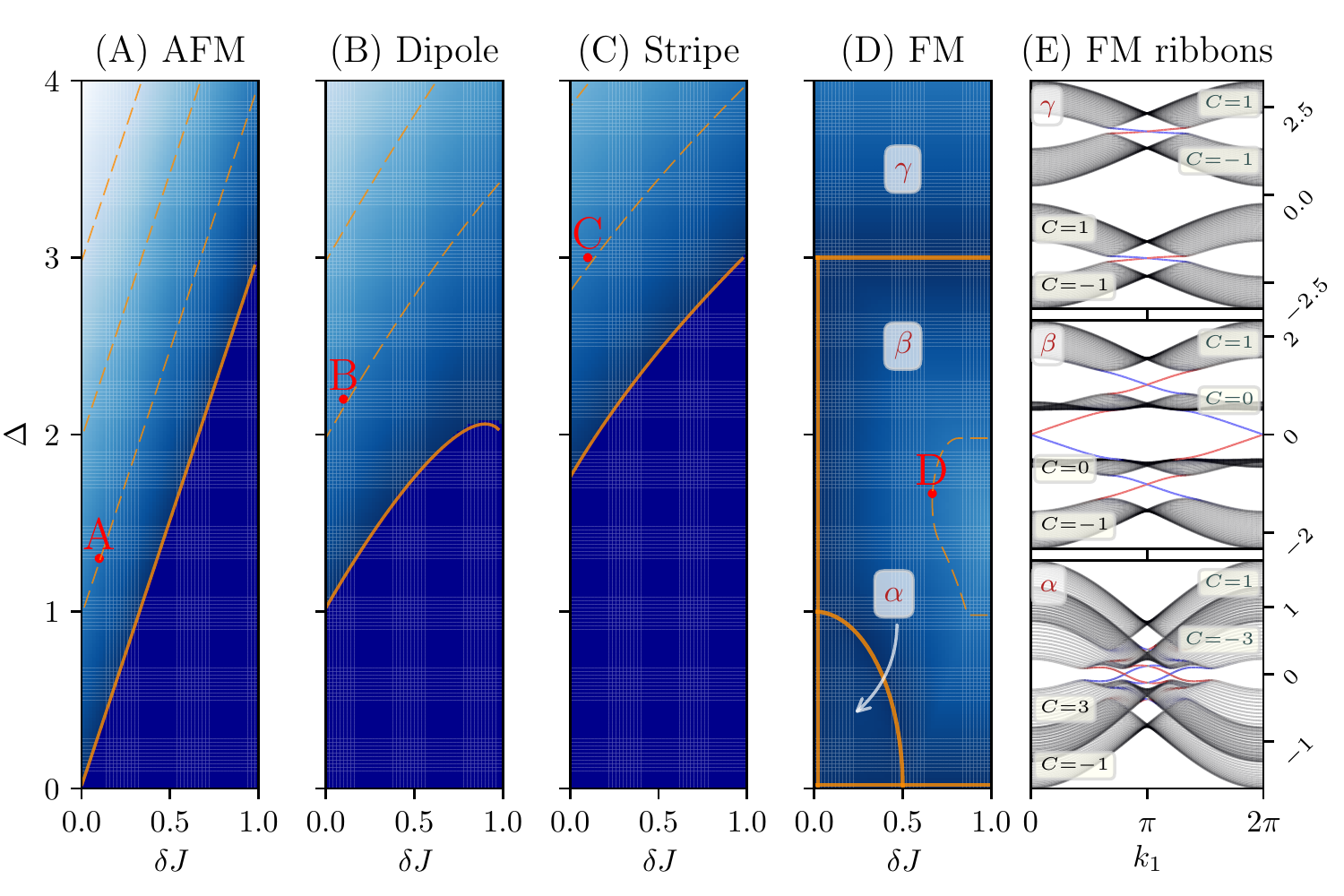}
  \caption{(A-D) Color maps of the energy gap $E_g$, for four spin phases, as a function of $\delta J$ and $\Delta$. The solid orange line separates the gapless (solid dark blue) and gapped (linearly fading blue) domains. The dashed orange lines are positive-integer-$E_g$ contours. (E) The band structures of FM-type ribbons for $(\alpha)$: $\delta J {=} 0.2$, $\Delta  {=} 0.4$, $(\beta)$: $\delta J {=} 2/3$, $\Delta {=} 5/3$, and $(\gamma)$: $\Delta J {=} 0.5$, $\Delta {=} 3.5$. The bands Chern numbers are displayed in light yellow boxes. Here $J{=}1$ is taken as a unit of measurement. Red letters A-D correspond to Figs. 2(a) to 2(d), respectively.}
  \label{fig:Egs_d&s}
\end{figure}

Since FM phase corresponds to the case of a uniform, out-of-plane, external magnetic field, a gap opens (see Fig.~\ref{fig:AFM_BS}d) between the bands, characterized by different Chern numbers, and bridged by chiral edge states~\cite{Nalitov-Z}. Fig.~\ref{fig:AFM_BS}d(i) displays the band structure of this phase, which is numerically obtained for $\delta J=2/3$, $\Delta=5/3$, with energy counted in units of $J$. We used the convention for which $K$ and $K'$ points in the first Brillouin zone are positioned at ($k_1, k_2$) equal to ($2\pi/3, -2\pi/3$), and vice versa, $\Gamma$ is placed at the origin, and $M$ at ($\pi, \pi$). Fig.~\ref{fig:AFM_BS}d(ii) shows a slice of the band structure along $\Gamma \to M \to K \to \Gamma$ pathway shown by the green solid line at the bottom of  Fig.~\ref{fig:AFM_BS}d(i). Figure~\ref{fig:AFM_BS}d(iii) illustrates the band structure of the graphene stripe (not to be confused with the "stripe" phase in Fig.~\ref{fig1}b) with zigzag edges revealing the topologically protected edge states, marked by red and blue colored lines for each edge respectively.

For the AFM phase, it can be shown that a gap opens when $\Delta > 3\delta J$, the gap value being $E_g = 2 (\Delta - 3 \delta J)$~\cite{Ohadi2015}. The gapped spectrum in the AFM lattice is characterized by trivial topology (see Fig.~\ref{fig:AFM_BS}a) and no edge states connecting the bulk bands. In Fig.~\ref{fig1}a the dotted white area in the AFM phase corresponds to the bands touching, whereas the yellow area to the bands being gapped. Here, the boundary separating the two regimes in Fig.~\ref{fig1}1a is calculated for the case of strong TE-TM splitting $\delta J/J = 0.5$. We point out that that gapped-ungapped AFM boundary coincides with the $W_\text{bif}$ boundary (red curve in Fig.~\ref{fig1}a) when $\delta J \to 0$. The AFM edge states shown in Fig.~\ref{fig:AFM_BS}a(iii), separated from the bands and marked by red and blue colors, are doubly degenerate and are not topologically protected. Figures~\ref{fig:Egs_d&s}(a,d) show phase diagrams for the AFM and FM spin phases as a function of effective Zeeman splitting and TE-TM splitting. The dark orange line marks the boundary between gapless and gapped phases. The points A and D correspond to Fig.~\ref{fig:AFM_BS}a and d, respectively. Figure~\ref{fig:Egs_d&s}e shows the band structures of FM-phase ribbons and serves to illustrate the effect of Chern numbers of the bands (indicated in Fig.~\ref{fig:Egs_d&s}d) on the dispersion of the chiral edge states.

To examine the band structure of the stripe and dipole phases, the following $8{\times}8$ Hamiltonian should be constructed in reciprocal $(k_1, k_2)$ space:
\begin{equation}{\label{eq:App_Ham8x8}}
    H_\vec{k} = \frac12 \Jtot (\vec{k}) + \frac{\Delta}{2} \cdot \diag{(\mu_1, \mu_2, \mu_3, \mu_4)} \otimes \hat{\sigma}_z.
\end{equation}
Note, that a unit cell in these two cases differs from that corresponding to AFM and FM phases and should be constructed as a pair of graphene unit cells taken successively, with the translational vectors being $\vec{a}_1$ and $2\vec{a}_2$. The first term in Eq.~\eqref{eq:App_Ham8x8}, $\Jtot (\vec{k})$, represents polaritonic graphene with TE-TM splitting and is specified in the Supplementary material \cite{Supplementary2019}, the second term is responsible for the magnetic patterns: for the dipole phase one sets $\mu_1 = \mu_4 = 1$, $\mu_2 = \mu_3 = -1$, for the stripe phase $\mu_1 = \mu_2 = 1$, $\mu_3 = \mu_4 = -1$. Both phases are characterized by topologically trivial band structure and edge states localized within the bulk (see Fig.~\ref{fig:AFM_BS}b,c). Figure~\ref{fig:Egs_d&s}(b,c) shows phase diagram for dipole and stripe phases with points B and C corresponding to the Fig.~\ref{fig:AFM_BS}b,c. In the trivial case $\delta J = 0$ one arrives to the gap opening condition $\Delta > J$ and $\Delta > \sqrt{3}J$ for dipole and stripe spin phases respectively which is plotted in Fig.~\ref{fig1}a indicating that gap opening only takes place at higher condensate densities (i.e., higher excitation powers).

\begin{figure}[t!]
  \includegraphics[width=\columnwidth]{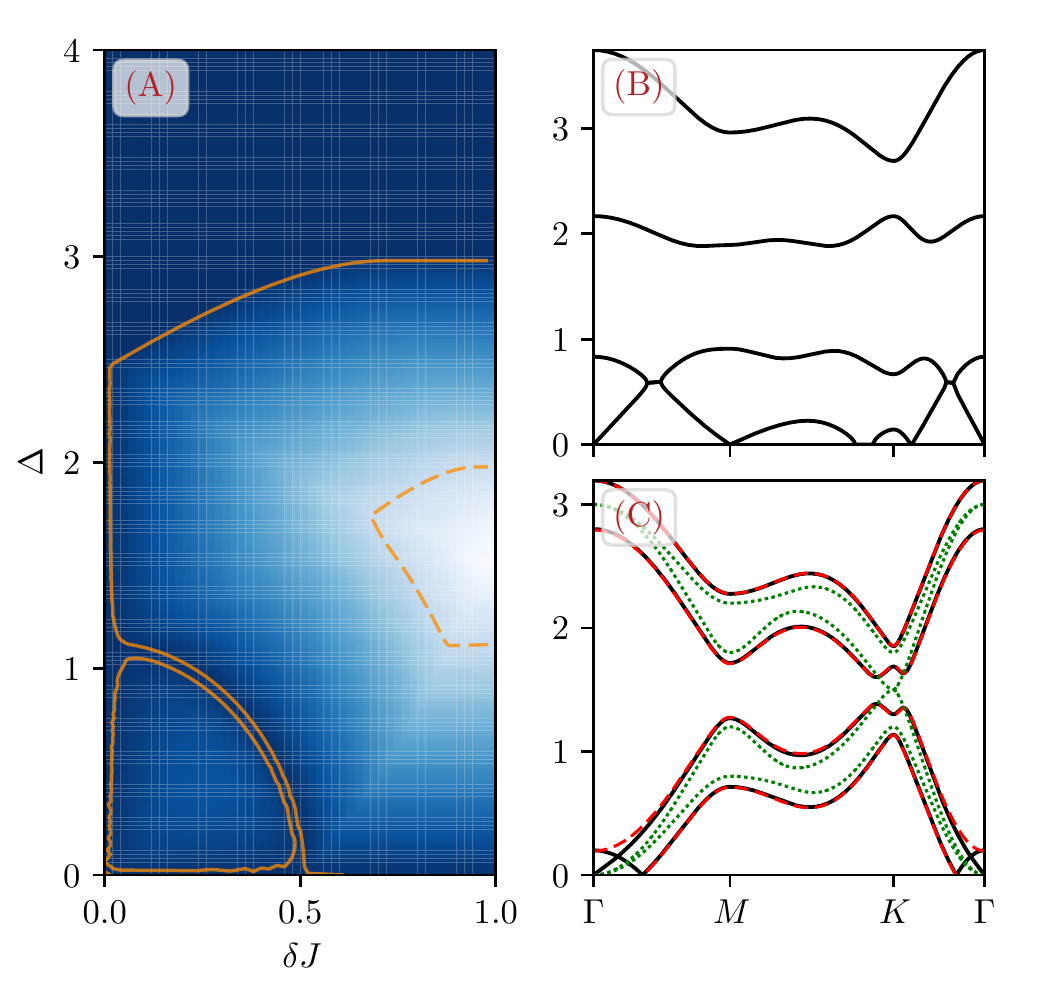}
  \caption{(A)~Colormap of the energy gap $E_g$ for the bogolons dispersions of the FM phase, as a function of $\delta J$ and $\Delta$. The solid orange line separates the gapless (solid dark blue) and gapped (linearly fading blue) domains. The dashed line is $E_g = J$ level contour. (B-C) The bogolons eigenenergies along $\Gamma \to M \to K \to \Gamma$ pathway for (B) $\delta J = 2J/3$, $\Delta = 5J/3$ and (C) $\delta J = 0.2J$, $\Delta = 0.4J$ (black solid lines). Additionally, green dotted and red dashed lines are the dispersions of linear model [Eq.~\eqref{eq:Hamiltonian4x4inK}] for $\Delta = 0$ and $\Delta = 0.4J$ respectively. $J=1$ is taken as a unit of measurement.}
  \label{fig:FM_bogolons}
\end{figure}

To investigate the FM phase in more detail, we performed Bogoliubov linearization of Eq.~\eqref{eq.GP}, generalized to all four spin patterns, and applied it to the FM one (see~[\onlinecite{Supplementary2019}] for more details). Figure~\ref{fig:FM_bogolons} demonstrates (A) the topological phase diagram for the effective spin-orbit interaction strength $\delta J / J$ and the effective interaction energy $\Delta$, and (B, C) eigenenergy curves along $\Gamma \to M \to K \to \Gamma$ pathway.
In addition, green dotted and red dashed curves show the dispersions of Eq.~\eqref{eq:Hamiltonian4x4inK} without and with effective magnetic field respectively, with the corresponding value $\Delta$.
Note that the bogolon dispersion overlaps with the one obtained in the effective magnetic field approximation in the region of high energies $E \gg \Delta$ and expectedly deviates for low energies (in the vicinity of $\Gamma$ point). This shows that the effective magnetic field model accurately describes the linear excitations of the condensate in the vicinity of the $K$-point.
Moreover, Fig.~\ref{fig:FM_bogolons}(A) clearly resembles Fig.~\ref{fig:Egs_d&s}(D).

\textit{Conclusions}. We have proposed an experimentally friendly geometry for realization of an optical $\mathbb{Z}$-topological insulator based on polaritonic graphene in the spin bifurcation regime. Differently from previous works, our proposal does not require application of an external magnetic field and the topological order appears spontaneously in the mean field picture through many-particle interactions under non-resonant and non-polarized pumping. The proposed effective field model due to interactions produces the topological phase diagram of the system and is consistent with the Bogoliubov model in a wide range of parameters.

\textit{Acknowledgements}. We thank Dr. O. Kyriienko for valuable discussions. The work was supported by Russian Science Foundation (Project No. 18-72-10110). A.V.N. and I.A.S. acknowledge support from Horizon2020 RISE project COEXAN. H.S. acknowledges support from UK’s Engineering and Physical Sciences Research Council (grant EP/M025330/1 on Hybrid Polaritonics). A.V.N. acknowledges support from Icelandic Research Fund, Grant No. 196301-051.


%%%%%%%%%%%%%%%%%%%%%%%%%%%%%%%%%%%%%%%%
\newpage
\renewcommand{\theequation}{S\arabic{equation}}
\renewcommand{\thefigure}{S\arabic{figure}}
\renewcommand{\thesection}{S\arabic{section}}

\begin{widetext}

\setcounter{equation}{0}
\setcounter{figure}{0}
\setcounter{table}{0}
\setcounter{page}{1}

\preprint{APS/123-QED}
\vspace{2cm}
\begin{center}
\textbf{\Large{Spontaneous topological transitions in polariton condensates due to spin bifurcations: Supplemental information}}
\end{center}

\section{Bogoliubov theory of spin-bifurcated polariton condensates in a honeycomb lattice}
We start with the following generalized Gross-Pitaevskii equation as given by Eq.~1 in main text. It describes the bulk honeycomb lattice of coherent spin-up ($+$) and spin-down ($-$) polaritons belonging to a condensate order parameter $\vec{\Psi} = (\psi_{1+},\psi_{1-},\psi_{2+},\psi_{2-}, \dots)^T$ where $T$ denotes the transpose:
\begin{equation}\label{eq:GP}
    i \frac{\partial \vec{\Psi}}{\partial t} =
 \left[
    \J + i (W-\Gamma)I_{2N} - (\epsilon + i\gamma) \, I_N \otimes \sigmax +
    \begin{pmatrix}
      \Shati{1} & & & \\
      & \Shati{2} & & \\
      & & \ddots & \\
      & & & \Shati{N}
    \end{pmatrix}
    - i\eta
    \begin{pmatrix}
      S_1 & & & \\
      & S_2 & & \\
      & & \ddots & \\
      & & & S_N
    \end{pmatrix}
    \otimes \sigmao
    \right]
    \vec{\Psi}.
\end{equation}
  Here we have omitted a factor $1/2$ appearing on the RHS in Eq.~1 in the main text for more transparent derivation. $\J$ incorporates all the hoppings between the sites of the lattice, $(W-\Gamma)$ is incoherent pumping-to-losses imbalance, $\epsilon + i\gamma$ is the complex splitting in the linear polarizations, $I_{2N}$ is the identity matrix of dimension $2N$ where $N$ is the number of condensates (sites) in the lattice which, for the case of $N=1$, we define $I_2 \equiv \sigmao$, and $\left\{ \sigmai{\lambda} \right\}_{\lambda=x, y, z}$ are the Pauli matrices. The ``spin'' operator $\Shati{A}$ is written,
  \begin{align}
    \Shati{A} = \alphaave S_A \sigmao + \alpha S_A^z \sigma,
  \end{align}
  where $\bar{\alpha} = \alpha_1 + \alpha_2$ and $\alpha = \alpha_1 - \alpha_2$ are  spin-anisotropic interaction parameters and
  \begin{equation} \label{eq.Sn}
S_A \equiv \frac{|\psi_{A+}|^2 + |\psi_{A-}|^2}{2}, \qquad S_{A}^z \equiv \frac{|\psi_{A+}|^2 - |\psi_{A-}|^2}{2}.
\end{equation}
  The block-diagonal $\Shati{}$- and $S$-operators are further designated as direct sums of the corresponding operator subspaces:
  \begin{align}
    &\begin{pmatrix}
      \Shati{1} & & & \\
      & \Shati{2} & & \\
      & & \ddots & \\
      & & & \Shati{N}
    \end{pmatrix} \equiv
    \bigoplus\limits_{A=1}^{N} \Shati{A} \quad \text{or simply} \quad \bigoplus\limits_A \Shati{A},
    \intertext{and}
    &\begin{pmatrix}
      S_1 & & & \\
      & S_2 & & \\
      & & \ddots & \\
      & & & S_N
    \end{pmatrix} \equiv
    \bigoplus\limits_{A} S_A.
  \end{align}
The Bogoliubov elementary excitation (bogolon) spectrum is obtained by the standard procedure of linearizing the Eq.~\eqref{eq:GP} around the steady state condensates $\vec{\Psi}_0 = \sqrt{\rho} \eo e^{-i \omega_0 t}$. We therefore seek $\vec{\Psi}$ in the form,
  \begin{equation}
    \vec{\Psi} = \left( \sqrt{\rho} \, \eo + \e^{-i\omega t} \vec{u} + \e^{i\omega^* t} \vec{v}^* \right) \e^{-i\omega_0 t},
  \end{equation}
  where the condensates density $\rho$ is taken to be uniform across the lattice, $\omega_0$ is the chemical potential, and $\eo$ captures their polarization,
  \begin{equation}{\label{eq:eo}}
    \eo =
    \begin{pmatrix}
      \vec{e}_1 \\
      \vec{e}_2 \\
      \vdots \\
      \vec{e}_N
    \end{pmatrix} =
    \begin{pmatrix}
      e_1^+ \\
      e_1^- \\
      e_2^+ \\
      e_2^- \\
      \vdots \\
      e_N^+ \\
      e_N^-
    \end{pmatrix} \equiv \bigoplus\limits_{j=1}^N \vec{e}_j.
  \end{equation}
Here, $\vec{u}$ and $\vec{v}$ are the amplitudes of the excitaions, of the same structure as $\eo$ \eqref{eq:eo}. Hereafter, $\e^{-i\omega t}$ and $\e^{i \omega^* t}$ are denoted as $\eplus$ and $\eminus$, respectively.

\section{Spins and $\Shati{}$-operators}
  For site $A$ we write,
\begin{subequations}
    \begin{align}
      S_A &= \rho S_{A0} + \rhohalf \left[
      \eplus \left(  \uvplus +  \uvminus  \right)
      +
      \eminus \left( \uvplusc   +  \uvminusc  \right)
      \right], \\
      S_A^z &= \rho S_{A0}^z + \rhohalf \left[
      \eplus \left( \uvplus  -  (\uvminus)  \right)
      +
      \eminus \left(  \uvplusc  -   (\uvminusc) \right)
      \right].
    \end{align}
\end{subequations}
We keep only linear order terms of $v_A^\pm, u_A^\pm$. $S_{A0}$ and $S_{A0}^z$ are defined as the normalized total spin and its $z$-projection corresponding to polarization $\ea$: $S_{A0} = \br{\abs{e_A^+}^2 + \abs{e_A^-}^2}/2$ and $S_{A0} = \br{\abs{e_A^+}^2 - \abs{e_A^-}^2}/2$.

  Consider $\Shati{}$-operators.
  \begin{subequations}
    \begin{align}
      \Shati{A} &= \alphaave S_A \sigmao + \alpha S_A^z \sigmaz =
      \rho \underbrace{\left( \alphaave S_{A0} \sigmao + \alpha S_{A0}^z \sigmaz \right)}_{\Shati{A0}} \nonumber \\
      & + \rhohalf 
      \begin{pmatrix}
        \boxed{\begin{aligned}
          &\eplus \left[ \substack{(\alphaave + \alpha) \br{\uvplus}  \\ + (\alphaave - \alpha) \br{\uvminus}} \right]  \\
          + &\eminus \left[ \substack{(\alphaave + \alpha) \br{\uvplusc}  \\ + (\alphaave - \alpha) \br{\uvminusc}} \right]
        \end{aligned}}
        & 0 \\
        0 &
        \boxed{\begin{aligned}
          &\eplus \left[ \substack{(\alphaave - \alpha) \br{\uvplus}  \\ + (\alphaave + \alpha) \br{\uvminus}} \right]  \\
          + &\eminus \left[ \substack{(\alphaave - \alpha) \br{\uvplusc}  \\ + (\alphaave + \alpha) \br{\uvminusc}} \right]
        \end{aligned}}
      \end{pmatrix}.
    \end{align}
    Further, we utilize equalities $\alphaave + \alpha = 2\alpha_1$, $\alphaave - \alpha = 2\alpha_2$. We assume that the condensates are fully spin polarized up $(|e_A^+|,|e_A^-|) = (1,0)$ or down $(|e_A^+|,|e_A^-|) = (0,1)$, and therefore $S_{A0} = 1/2$. For spin-up(down) polarization we will then have $S_{A0}^{z \uparrow} = 1/2$ and $S_{A0}^{z \downarrow} = -1/2$ respectively and
    \begin{equation}{\label{eq:SA0}}
      \Shati{A0} =
      \begin{pmatrix}
        \alphaave S_{A0} + \alpha S_{A0}^z & 0 \\
        0 & \alphaave S_{A0} - \alpha S_{A0}^z
      \end{pmatrix} =
      \left\{
      \begin{aligned}
        &\uparrow, \quad
        \begin{pmatrix}
          \alpha_1 & 0 \\
          0 & \alpha_2
        \end{pmatrix}, \\
        &\downarrow, \quad
        \begin{pmatrix}
          \alpha_2 & 0 \\
          0 & \alpha_1
        \end{pmatrix}.
      \end{aligned}
      \right.
    \end{equation}
    For the spin operators:
    \begin{align}
      \Shati{A} = \Shati{A0} + \sqrt{\rho}
      {%\tiny
      \begin{pmatrix}
        \boxed{\tiny
        \begin{aligned}
          &\eplus \left[ \alpha_1 \br{\uvplus} + \alpha_2 \br{\uvminus} \right] \\
          + &\eminus \left[ \alpha_1 \br{\uvplusc} + \alpha_2 \br{\uvminusc} \right]
        \end{aligned}
        }
        & 0 \\
        0 &
        \boxed{\tiny
        \begin{aligned}
          &\eplus \left[ \alpha_2 \br{\uvplus} + \alpha_1 \br{\uvminus} \right] \\
          + &\eminus \left[ \alpha_2 \br{\uvplusc} + \alpha_1 \br{\uvminusc} \right]
        \end{aligned}
        }
      \end{pmatrix}}
      \label{eq:Shat_prefin}
    \end{align}
    Next, as we finally need to equate coefficients of $\eplus$ ($\eminus$) on both the sides of \eqref{eq:GP}, we rearrange the terms in \eqref{eq:Shat_prefin}:
    \begin{align}
      \Shati{A} = \Shati{A0} +
      &\sqrt{\rho} \, \eplus
      \underbrace{\tiny
      \begin{pmatrix}
        \alpha_1 \br{\uvplus} + \alpha_2 \br{\uvminus} & 0 \\
        0 & \alpha_2 \br{\uvplus} + \alpha_1 \br{\uvminus}
      \end{pmatrix}}_{\Shati{A+}} \\
      + &\sqrt{\rho} \, \eminus
      \underbrace{{\tiny
      \begin{pmatrix}
        \phantom{\alpha_1 \br{\uvplus} + \alpha_2 \br{\uvminus}} & \\
        & \phantom{\alpha_2 \br{\uvplus} + \alpha_1 \br{\uvminus}}
      \end{pmatrix}}^*}_{\Shati{A-} = \Shati{A+}^*}
      \\
      &= \Shati{A0} + \Shati{A+} + \Shati{A-}.
    \end{align}
  \end{subequations}

  \section{Hamiltonian with respect to $\eplus$ and $\eminus$}
We now decompose Eq.~\eqref{eq:GP}, separating the terms with $\eplus$ and $\eminus$ multipliers.
  \begin{subequations}
    \begin{align}
      \label{eq:H0}
      \Htot &= \overbrace{\J + i (W-\Gamma)I_{2N} - (\epsilon + i\gamma) \, I_N \otimes \sigmax + \rho \bigoplus\limits_A \Shati{A0} - i\eta \rho \underbrace{\bigoplus\limits_A S_{A0}}_{I_{2N}/2}}^{\Hzero} \\
      &+ \eplus \, \sqrt{\rho}
      \left[
      \bigoplus\limits_A \Shati{A+} - \frac{i\eta}{2} \bigoplus\limits_A \left( \uvplus + \uvminus \right) \otimes \sigmao
      \right] \\
      &+ \eminus \, \sqrt{\rho}
      \left[
      \bigoplus\limits_A \Shati{A-} - \frac{i\eta}{2} \bigoplus\limits_A \left( \uvplusc + \uvminusc \right) \otimes \sigmao
      \right]
    \end{align}
    \begin{align}
      \label{eq:Splus_def}
      = \Hzero &+ \eplus \sqrt{\rho} \ \underbrace{\bigoplus\limits_A
      {\footnotesize
      \begin{pmatrix}
        \boxed{
        \begin{aligned}
          &\br{\uvplus} \br{\alpha_1 - \frac{i\eta}{2}} \\
          + &\br{\uvminus} \br{\alpha_2 - \frac{i\eta}{2}}
        \end{aligned}} & 0 \\
        0 &
        \boxed{
        \begin{aligned}
          &\br{\uvplus} \br{\alpha_2 - \frac{i\eta}{2}} \\
          + &\br{\uvminus} \br{\alpha_1 - \frac{i\eta}{2}}
        \end{aligned}}
      \end{pmatrix}}}_{\Splus (\eta)} \\
      &+ \eminus \sqrt{\rho} \ \Splus^*(-\eta) = \Hzero + \Splus(\eta) + \Sminus(\eta).
    \end{align}
  \end{subequations}
  Then, we equate the corresponding coefficients in front of $\eplus$ and $\eminus$:
  \begin{equation}
    \label{eq:omega_uv_prefin}
    \left\{
    \begin{aligned}
      \omega \vec{u} &= \phantom{-} \left(\Hzero - \omega_0 I_{2N} \right) \vec{u} + \rho \, \Splus \, \eo, \\
      \omega \vec{v} &= - \left( \Hzero^* - \omega_0 I_{2N} \right) \vec{v} - \rho \, \Sminus^* \, \eo^*.
    \end{aligned}
    \right.
  \end{equation}
  Further we reveal the action of operator $\Splus$ to $\eo$, as it contains $\vec{u}$ and $\vec{v}$ terms.

As we consider totally polarized condensate, $\ea = (\e^{i\phi_A}, 0)^T$ or $(0, \e^{i\phi_A})^T$ with unknown phase $\phi_A$ we satisfy the realtion $e_A^+ \, e_A^- = 0$. From Eq.~\eqref{eq:eo} and \eqref{eq:Splus_def}, we get after some algebra
  \begin{subequations}
    \label{eq:S+e0_S-e0}
    \begin{align}
      \nonumber
      \Splus \, \eo &=
      \left[ \bigoplus\limits_A
      {\scriptsize
      \begin{pmatrix}
        \boxed{
        \begin{aligned}
          &\br{\uvplus} \br{\alpha_1 - \frac{i\eta}{2}} \\
          + &\br{\uvminus} \br{\alpha_2 - \frac{i\eta}{2}}
        \end{aligned}} & 0 \\
        0 &
        \boxed{
        \begin{aligned}
          &\br{\uvplus} \br{\alpha_2 - \frac{i\eta}{2}} \\
          + &\br{\uvminus} \br{\alpha_1 - \frac{i\eta}{2}}
        \end{aligned}}
      \end{pmatrix}}
      \right]
      \
      \left[ \bigoplus\limits_A \ea \right] \\
      &= \bigoplus\limits_A
      {\scriptsize
      \begin{pmatrix}
        \br{\alphaetaone} \br{\uvplus} e_A^+ +
        \br{\alphaetatwo} \cancelto{0}{\br{\uvminus} e_A^+} \\
        \br{\alphaetatwo} \cancelto{0}{\br{\uvplus} e_A^-} +
        \br{\alphaetaone} \br{\uvminus} e_A^-
      \end{pmatrix}
      }
    \end{align}
    \begin{align}
      \\
      &=
      \br{\alphaetaone} \cdot \left\{
      \underbrace{
      \left[ \bigoplus\limits_A
      \begin{pmatrix}
        \abs{e_A^+}^2 & 0 \\
        0 & \abs{e_A^-}^2
      \end{pmatrix}
      \right]
      }_{\Ezero'}
      \vec{u}
      +
      \underbrace{
      \left[ \bigoplus\limits_A
      \begin{pmatrix}
        \br{e_A^+}^2 & 0 \\
        0 & \br{e_A^-}^2
      \end{pmatrix}
      \right]
      }_{\Ezero}
      \vec{v} \right\}.
      \label{eq:Splus_final}
    \end{align}
    Analogously, we can obtain
    \begin{equation}
      \Sminus^* \, \eo^* = \Splus(-\eta) \, \eo^* = \br{\alphaetaoneplus} \cdot
      \left\{ \Ezero^* \vec{u} + \Ezero' \vec{v} \right\}.
    \end{equation}
  \end{subequations}
  Inserting \eqref{eq:S+e0_S-e0} into \eqref{eq:omega_uv_prefin}, we arrive at
  \begin{equation}
    \left\{
    \begin{aligned}
      \omega \vec{u} &= \phantom{-} \br{\Hzero - \omega_0 I_{2N}} \vec{u} + \rho \br{\alphaetaone} \br{\Ezero' \vec{u} + \Ezero \vec{v}}, \\
      \omega \vec{v} &= - \br{\Hzero^* - \omega_0 I_{2N}} \vec{v} - \rho \br{\alphaetaoneplus} \br{\Ezero^* \vec{u} + \Ezero' \vec{v}},
    \end{aligned}
    \right.
  \end{equation}
  and finally
  \begin{equation}
    \label{eq:final_res}
    \omega
    \begin{pmatrix}
      \vec{u} \\
      \vec{v}
    \end{pmatrix} =
    \begin{pmatrix}
      \Hzero - \omega_0 I_{2N} - \rho \left( \frac{i\eta}{2} - \alpha_1 \right) \Ezero' &
      - \rho \br{\frac{i\eta}{2} - \alpha_1} \Ezero \\
      - \rho \br{\frac{i\eta}{2} + \alpha_1} \Ezero^* &
      - \br{\Hzero^* - \omega_0 I_{2N}} - \rho \br{\frac{i\eta}{2} + \alpha_1} \Ezero'
    \end{pmatrix}
    \begin{pmatrix}
      \vec{u} \\
      \vec{v}
    \end{pmatrix}.
  \end{equation}
  Each position in the matrix \eqref{eq:final_res} is of $(2N{\times}2N)$ dimension, so the final matrix to be diagonalized is of size $(4N{\times}4N)$.

\section{Extracting $\rho$ and $\omega_0$ for the stationary condensates}
Prior to the diagonalization of Eq.~\eqref{eq:final_res}, we should obtain quantities $\omega_0$ and $\rho$. To find them, we use the equality
  \begin{equation}
    \omega_0 = \frac{\langle \vec{\Psi}_0 | \hat{H}_0 | \vec{\Psi}_0 \rangle}{\langle \vec{\Psi}_0 | \vec{\Psi}_0 \rangle},
  \end{equation}
where $\Htot_0$ is defined in Eq.~\eqref{eq:H0}. Then,
  \begin{subequations}
    \begin{align}
      \omega_0 \, N &= \langle \eo | \J + i(W-\Gamma-\eta\rho/2) \, I_{2N} - (\epsilon + i\gamma) \, I_N \otimes \sigmax + \rho\bigoplus\limits_A \Shati{A0} | \eo \rangle \\
      &= \langle \eo | \J | \eo \rangle + i(W-\Gamma-\eta\rho/2) \, N -
      (\epsilon + i\gamma) \langle \eo | I_N \otimes \sigmax | \eo \rangle +
      \rho \langle \eo | \bigoplus\limits_A \Shati{A0} | \eo \rangle
      \label{eq:omega0_1}
    \end{align}
    We can neglect the third term in \eqref{eq:omega0_1} for the strongly polarized condensates. Consider the last term:
    \begin{align}
      \langle \ea | \Shati{A0} | \ea \rangle =
      \langle \ea | \alphaave S_{A0} \sigmao + \alpha S_{A0}^z \sigmaz | \ea \rangle =
      2 \alphaave S_{A0}^2 + 2 \alpha \left( S_{A0}^{z} \right)^2 = \frac{\alphaave + \alpha}{2} = \alpha_1.
    \end{align}
    Hence,
    \begin{equation}
      \label{eq:omega0_2}
      \omega_0 = \frac{\langle \eo | \J | \eo \rangle + i(W-\Gamma-\eta\rho/2) \, N + \rho \, N \, \alpha_1}{N}.
    \end{equation}
  \end{subequations}
  The density $\rho$ is derived from the requirement of the real-value essence of $\omega_0$:
  \begin{equation}\label{eq:rho_fin}
    \rho = \frac{2 \, (W - \Gamma)}{\eta}.
  \end{equation}
  Thus,
  \begin{equation}\label{eq:omega0_fin}
    \omega_0 = \frac{\langle \eo | \J | \eo \rangle}{N} + \rho \, \alpha_1.
  \end{equation}
  For $\delta J \approx 0$, the first term can be evaluated due to $\langle \eo | \J | \eo \rangle = -3NJ, -2NJ, -NJ$, and $0$ for the ferromagnetic, stripe, dipole, and antiferromagnetic phases, respectively.

  \section{Band structures of the bogolons}
  In order to get the bogolon dispersion, we transform the external spatial degrees of freedom into the lattice momentum $\mathbf{k}$ (reciprocal) basis. In Eq.~\eqref{eq:final_res}, we substitute $\vec{u}$ and $\vec{v}$ with separated intra-cell and inter-cell degrees of freedom and Fourier-transform along the inter-cell coordinate. The bulk Hamiltonian is then written,
  \begin{equation}\label{eq:Hbulk}
    \Hbulk(\vec{k}) =
    \begin{pmatrix}
      \Hzero(\vec{k}) - \omega_0 I_{2\ns} - \rho \left( \frac{i\eta}{2} - \alpha_1 \right) \Ezero'(\ns) &
      - \rho \br{\frac{i\eta}{2} - \alpha_1} \Ezero(\ns) \\
      - \rho \br{\frac{i\eta}{2} + \alpha_1} \Ezero^*(\ns) &
      - \br{\Hzero^*(-\vec{k}) - \omega_0 I_{2\ns}} - \rho \br{\frac{i\eta}{2} + \alpha_1} \Ezero'(\ns)
    \end{pmatrix}
  \end{equation}
  The number of sites per unit cell is denoted as $\ns$. $\Ezero$, $\Ezero'$, and $\Ezero^*$ with $\ns$ positioned as an argument are taken over a single unit cell, i.e. with the summation $\bigoplus\limits_A$ limited to sites $A$ within a single unit cell.

Substituting the density \eqref{eq:rho_fin} into $\Hzero$, we arrive at
  \begin{equation}
    \Hzero (\vec{k}) = \J (\vec{k}) - (\epsilon + i\gamma) I_N \otimes \sigmax + \rho \bigoplus\limits_A \Shati{A0}.
  \end{equation}
  To present the explicit form of the bulk Hamiltonian \eqref{eq:Hbulk}, we denote the hopping $2{\times}2$ blocks by $\jhat{i}$:
  \begin{equation}
    \jhat{i} = -
    \begin{pmatrix}
      J & \delta J \, \e^{-2i\varphi_i} \\
      \delta J \, \e^{2i\varphi} & J
    \end{pmatrix}
  \end{equation}
  with the set of angles $\varphi$ chosen as $(0, 2\pi/3, 4\pi/3)$. Then,
  \begin{equation}{\label{eq:Jhopk}}
    \J(k_1, k_2) =
    \begin{pmatrix}
      \O & \jhat{1} + \jhat{3} \, \e^{-ik_1} & \O & \jhat{2} \, \e^{-ik_2} \\
      \jhat{1} + \jhat{3} \, \e^{ik_1} & \O & \jhat{2} & \O \\
      \O & \jhat{2} & \O & \jhat{1} + \jhat{3} \, \e^{-ik_1} \\
      \jhat{2} \, \e^{ik_2} & \O & \jhat{1} + \jhat{3} \, \e^{ik_1} & \O
    \end{pmatrix}.
  \end{equation}
  where $\O$ is the $2\times2$ zero matrix. The blocks $\Shati{A0}$ for each distinct polarization pattern belonging to the spin-bifurcated condensates (see Fig.~1b in main text) are written:
  \begin{align*}
    \Shati{A0,\text{AFM}} &=
    {\scriptsize
    \begin{pmatrix}
      \alpha_1 & & & & & & & \\
      & \alpha_2 & & & & & & \\
      & & \alpha_2 & & & & & \\
      & & & \alpha_1 & & & & \\
      & & & & \alpha_1 & & & \\
      & & & & & \alpha_2 & & \\
      & & & & & & \alpha_2 & \\
      & & & & & & & \alpha_1
    \end{pmatrix}}, \
   &  \Shati{A0,\text{Dipole}} =
    {\scriptsize
    \begin{pmatrix}
      \alpha_1 & & & & & & & \\
      & \alpha_2 & & & & & & \\
      & & \alpha_2 & & & & & \\
      & & & \alpha_1 & & & & \\
      & & & & \alpha_2 & & & \\
      & & & & & \alpha_1 & & \\
      & & & & & & \alpha_1 & \\
      & & & & & & & \alpha_2
    \end{pmatrix}}, \\
    \Shati{A0,\text{Stripe}} &=
    {\scriptsize
    \begin{pmatrix}
      \alpha_1 & & & & & & & \\
      & \alpha_2 & & & & & & \\
      & & \alpha_1 & & & & & \\
      & & & \alpha_2 & & & & \\
      & & & & \alpha_2 & & & \\
      & & & & & \alpha_1 & & \\
      & & & & & & \alpha_2 & \\
      & & & & & & & \alpha_1
    \end{pmatrix}}, \
    & \Shati{A0,\text{FM}} =
    {\scriptsize
    \begin{pmatrix}
      \alpha_1 & & & & & & & \\
      & \alpha_2 & & & & & & \\
      & & \alpha_1 & & & & & \\
      & & & \alpha_2 & & & & \\
      & & & & \alpha_1 & & & \\
      & & & & & \alpha_2 & & \\
      & & & & & & \alpha_1 & \\
      & & & & & & & \alpha_2
    \end{pmatrix}}.
  \end{align*}

\section{Connection with the model of effective magnetic field}
We will consider the scenario where the in-plane magnetic field, caused by birefringence in the cavity and/or its distributed-Bragg-reflectors, is much smaller than the condensate induced out-of-plane magnetic field, i.e. $\alpha \rho \gg \epsilon$ in frequency units. This can be achieved by considering Eq.~\eqref{eq:rho_fin} which dictates that $\rho$ can be increased by simply tuning the ratio of pump-decay ($W-\Gamma$) against the saturation rate $(\eta)$. If the saturation rate $\eta$ is taken to be smaller than then same-spin polariton-polariton interaction strength $\alpha_1$ then we have,
  \begin{equation}{\label{eq:toy_Hbulk1}}
    \Hbulk (\vec{k}) = \frac12
    {\footnotesize
    \begin{pmatrix}
      \J(\vec{k}) + \rho\bigoplus\limits_A \Shati{A0} - \omega_0 I_{2\ns} + \rho\alpha_1 \Ezero'(\ns)
      & \rho \alpha_1 \Ezero(\ns) \\
      -\rho\alpha_1 \Ezero^*(\ns)
      & -\J^* (-\vec{k}) - \rho\bigoplus\limits_A \Shati{A0} + \omega_0 I_{2\ns} - \rho\alpha_1 \Ezero'(\ns)
    \end{pmatrix}
    }
  \end{equation}
  The prefactor $1/2$ originally omitted from Eq.~\eqref{eq:GP} is now returned back. Equation~\eqref{eq:toy_Hbulk1} then describes the spinor bogolon dispersion experiencing an effective out-of-plane magnetic field. Results from diagonalizing the above Hamiltonian are presented in the main text (Fig.~4).

In the linear regime ($\rho \simeq 0$) there is no effective out-of-plane magnetic field and the bogolon dispersions become analogous to the dispersions of $\J(\vec{k}) / 2$ graphene Hamiltonian. The effective-magnetic-field model dispersions presented in Figures~2 and 3 in the main text are then the result of diagonalization of
  \begin{equation}
    H_{\text{e.m.f.}} (\vec{k}) = \frac12 \left( \J(\vec{k}) + \rho \bigoplus\limits_A \Shati{A0} \right).
  \end{equation}
  Within the notations, the magnitude of the magnetic field is $\rho\alpha/2$, in accordance with the definition of $\Delta$ in the main text.

\end{widetext}


\begin{thebibliography}{38}%
\makeatletter
\providecommand \@ifxundefined [1]{%
 \@ifx{#1\undefined}
}%
\providecommand \@ifnum [1]{%
 \ifnum #1\expandafter \@firstoftwo
 \else \expandafter \@secondoftwo
 \fi
}%
\providecommand \@ifx [1]{%
 \ifx #1\expandafter \@firstoftwo
 \else \expandafter \@secondoftwo
 \fi
}%
\providecommand \natexlab [1]{#1}%
\providecommand \enquote  [1]{``#1''}%
\providecommand \bibnamefont  [1]{#1}%
\providecommand \bibfnamefont [1]{#1}%
\providecommand \citenamefont [1]{#1}%
\providecommand \href@noop [0]{\@secondoftwo}%
\providecommand \href [0]{\begingroup \@sanitize@url \@href}%
\providecommand \@href[1]{\@@startlink{#1}\@@href}%
\providecommand \@@href[1]{\endgroup#1\@@endlink}%
\providecommand \@sanitize@url [0]{\catcode `\\12\catcode `\$12\catcode
  `\&12\catcode `\#12\catcode `\^12\catcode `\_12\catcode `\%12\relax}%
\providecommand \@@startlink[1]{}%
\providecommand \@@endlink[0]{}%
\providecommand \url  [0]{\begingroup\@sanitize@url \@url }%
\providecommand \@url [1]{\endgroup\@href {#1}{\urlprefix }}%
\providecommand \urlprefix  [0]{URL }%
\providecommand \Eprint [0]{\href }%
\providecommand \doibase [0]{http://dx.doi.org/}%
\providecommand \selectlanguage [0]{\@gobble}%
\providecommand \bibinfo  [0]{\@secondoftwo}%
\providecommand \bibfield  [0]{\@secondoftwo}%
\providecommand \translation [1]{[#1]}%
\providecommand \BibitemOpen [0]{}%
\providecommand \bibitemStop [0]{}%
\providecommand \bibitemNoStop [0]{.\EOS\space}%
\providecommand \EOS [0]{\spacefactor3000\relax}%
\providecommand \BibitemShut  [1]{\csname bibitem#1\endcsname}%
\let\auto@bib@innerbib\@empty
%</preamble>
\bibitem [{\citenamefont {Hasan}\ and\ \citenamefont {Kane}(2010)}]{Hasan2010}%
  \BibitemOpen
  \bibfield  {author} {\bibinfo {author} {\bibfnamefont {M.~Z.}\ \bibnamefont
  {Hasan}}\ and\ \bibinfo {author} {\bibfnamefont {C.~L.}\ \bibnamefont
  {Kane}},\ }\href {\doibase 10.1103/RevModPhys.82.3045} {\bibfield  {journal}
  {\bibinfo  {journal} {Rev. Mod. Phys.}\ }\textbf {\bibinfo {volume} {82}},\
  \bibinfo {pages} {3045} (\bibinfo {year} {2010})}\BibitemShut {NoStop}%
\bibitem [{\citenamefont {Qi}\ and\ \citenamefont {Zhang}(2011)}]{Qi2011}%
  \BibitemOpen
  \bibfield  {author} {\bibinfo {author} {\bibfnamefont {X.-L.}\ \bibnamefont
  {Qi}}\ and\ \bibinfo {author} {\bibfnamefont {S.-C.}\ \bibnamefont {Zhang}},\
  }\href {\doibase 10.1103/RevModPhys.83.1057} {\bibfield  {journal} {\bibinfo
  {journal} {Rev. Mod. Phys.}\ }\textbf {\bibinfo {volume} {83}},\ \bibinfo
  {pages} {1057} (\bibinfo {year} {2011})}\BibitemShut {NoStop}%
\bibitem [{\citenamefont {Bansil}\ \emph {et~al.}(2016)\citenamefont {Bansil},
  \citenamefont {Lin},\ and\ \citenamefont {Das}}]{Bansil2016}%
  \BibitemOpen
  \bibfield  {author} {\bibinfo {author} {\bibfnamefont {A.}~\bibnamefont
  {Bansil}}, \bibinfo {author} {\bibfnamefont {H.}~\bibnamefont {Lin}}, \ and\
  \bibinfo {author} {\bibfnamefont {T.}~\bibnamefont {Das}},\ }\href {\doibase
  10.1103/RevModPhys.88.021004} {\bibfield  {journal} {\bibinfo  {journal}
  {Rev. Mod. Phys.}\ }\textbf {\bibinfo {volume} {88}},\ \bibinfo {pages}
  {021004} (\bibinfo {year} {2016})}\BibitemShut {NoStop}%
\bibitem [{\citenamefont {Kane}\ and\ \citenamefont {Mele}(2005)}]{Kane2005a}%
  \BibitemOpen
  \bibfield  {author} {\bibinfo {author} {\bibfnamefont {C.~L.}\ \bibnamefont
  {Kane}}\ and\ \bibinfo {author} {\bibfnamefont {E.~J.}\ \bibnamefont
  {Mele}},\ }\href {\doibase 10.1103/PhysRevLett.95.226801} {\bibfield
  {journal} {\bibinfo  {journal} {Phys. Rev. Lett.}\ }\textbf {\bibinfo
  {volume} {95}},\ \bibinfo {pages} {226801} (\bibinfo {year}
  {2005})}\BibitemShut {NoStop}%
\bibitem [{\citenamefont {Fu}\ and\ \citenamefont {Kane}(2007)}]{Fu2007}%
  \BibitemOpen
  \bibfield  {author} {\bibinfo {author} {\bibfnamefont {L.}~\bibnamefont
  {Fu}}\ and\ \bibinfo {author} {\bibfnamefont {C.~L.}\ \bibnamefont {Kane}},\
  }\href {\doibase 10.1103/PhysRevB.76.045302} {\bibfield  {journal} {\bibinfo
  {journal} {Physical Review B}\ }\textbf {\bibinfo {volume} {76}},\ \bibinfo
  {pages} {045302} (\bibinfo {year} {2007})}\BibitemShut {NoStop}%
\bibitem [{\citenamefont {Moore}\ and\ \citenamefont
  {Balents}(2007)}]{Moore2007}%
  \BibitemOpen
  \bibfield  {author} {\bibinfo {author} {\bibfnamefont {J.~E.}\ \bibnamefont
  {Moore}}\ and\ \bibinfo {author} {\bibfnamefont {L.}~\bibnamefont
  {Balents}},\ }\href {\doibase 10.1103/PhysRevB.75.121306} {\bibfield
  {journal} {\bibinfo  {journal} {Phys. Rev. B}\ }\textbf {\bibinfo {volume}
  {75}},\ \bibinfo {pages} {121306(R)} (\bibinfo {year} {2007})}\BibitemShut
  {NoStop}%
\bibitem [{\citenamefont {Roy}(2009)}]{Roy2009}%
  \BibitemOpen
  \bibfield  {author} {\bibinfo {author} {\bibfnamefont {R.}~\bibnamefont
  {Roy}},\ }\href {\doibase 10.1103/PhysRevB.79.195322} {\bibfield  {journal}
  {\bibinfo  {journal} {Phys. Rev. B}\ }\textbf {\bibinfo {volume} {79}},\
  \bibinfo {pages} {195322} (\bibinfo {year} {2009})}\BibitemShut {NoStop}%
\bibitem [{\citenamefont {Xia}\ \emph {et~al.}(2009)\citenamefont {Xia},
  \citenamefont {Qian}, \citenamefont {Hsieh}, \citenamefont {Wray},
  \citenamefont {Pal}, \citenamefont {Lin}, \citenamefont {Bansil},
  \citenamefont {Grauer}, \citenamefont {Hor}, \citenamefont {Cava},\ and\
  \citenamefont {Hasan}}]{Xia2009}%
  \BibitemOpen
  \bibfield  {author} {\bibinfo {author} {\bibfnamefont {Y.}~\bibnamefont
  {Xia}}, \bibinfo {author} {\bibfnamefont {D.}~\bibnamefont {Qian}}, \bibinfo
  {author} {\bibfnamefont {D.}~\bibnamefont {Hsieh}}, \bibinfo {author}
  {\bibfnamefont {L.}~\bibnamefont {Wray}}, \bibinfo {author} {\bibfnamefont
  {A.}~\bibnamefont {Pal}}, \bibinfo {author} {\bibfnamefont {H.}~\bibnamefont
  {Lin}}, \bibinfo {author} {\bibfnamefont {A.}~\bibnamefont {Bansil}},
  \bibinfo {author} {\bibfnamefont {D.}~\bibnamefont {Grauer}}, \bibinfo
  {author} {\bibfnamefont {Y.~S.}\ \bibnamefont {Hor}}, \bibinfo {author}
  {\bibfnamefont {R.~J.}\ \bibnamefont {Cava}}, \ and\ \bibinfo {author}
  {\bibfnamefont {M.~Z.}\ \bibnamefont {Hasan}},\ }\href {\doibase
  10.1038/nphys1274} {\bibfield  {journal} {\bibinfo  {journal} {Nature
  Physics}\ }\textbf {\bibinfo {volume} {5}},\ \bibinfo {pages} {398} (\bibinfo
  {year} {2009})}\BibitemShut {NoStop}%
\bibitem [{\citenamefont {Zhang}\ \emph {et~al.}(2009)\citenamefont {Zhang},
  \citenamefont {Liu}, \citenamefont {Qi}, \citenamefont {Dai}, \citenamefont
  {Fang},\ and\ \citenamefont {Zhang}}]{Zhang2009}%
  \BibitemOpen
  \bibfield  {author} {\bibinfo {author} {\bibfnamefont {H.}~\bibnamefont
  {Zhang}}, \bibinfo {author} {\bibfnamefont {C.-X.}\ \bibnamefont {Liu}},
  \bibinfo {author} {\bibfnamefont {X.-L.}\ \bibnamefont {Qi}}, \bibinfo
  {author} {\bibfnamefont {X.}~\bibnamefont {Dai}}, \bibinfo {author}
  {\bibfnamefont {Z.}~\bibnamefont {Fang}}, \ and\ \bibinfo {author}
  {\bibfnamefont {S.-C.}\ \bibnamefont {Zhang}},\ }\href {\doibase
  10.1038/nphys1270} {\bibfield  {journal} {\bibinfo  {journal} {Nature
  Physics}\ }\textbf {\bibinfo {volume} {5}},\ \bibinfo {pages} {438} (\bibinfo
  {year} {2009})}\BibitemShut {NoStop}%
\bibitem [{\citenamefont {Chen}\ \emph {et~al.}(2009)\citenamefont {Chen},
  \citenamefont {Analytis}, \citenamefont {Chu}, \citenamefont {Liu},
  \citenamefont {Mo}, \citenamefont {Qi}, \citenamefont {Zhang}, \citenamefont
  {Lu}, \citenamefont {Dai}, \citenamefont {Fang}, \citenamefont {Zhang},
  \citenamefont {Fisher}, \citenamefont {Hussain},\ and\ \citenamefont
  {Shen}}]{Chen2009}%
  \BibitemOpen
  \bibfield  {author} {\bibinfo {author} {\bibfnamefont {Y.~L.}\ \bibnamefont
  {Chen}}, \bibinfo {author} {\bibfnamefont {J.~G.}\ \bibnamefont {Analytis}},
  \bibinfo {author} {\bibfnamefont {J.-H.}\ \bibnamefont {Chu}}, \bibinfo
  {author} {\bibfnamefont {Z.~K.}\ \bibnamefont {Liu}}, \bibinfo {author}
  {\bibfnamefont {S.-K.}\ \bibnamefont {Mo}}, \bibinfo {author} {\bibfnamefont
  {X.~L.}\ \bibnamefont {Qi}}, \bibinfo {author} {\bibfnamefont {H.~J.}\
  \bibnamefont {Zhang}}, \bibinfo {author} {\bibfnamefont {D.~H.}\ \bibnamefont
  {Lu}}, \bibinfo {author} {\bibfnamefont {X.}~\bibnamefont {Dai}}, \bibinfo
  {author} {\bibfnamefont {Z.}~\bibnamefont {Fang}}, \bibinfo {author}
  {\bibfnamefont {S.~C.}\ \bibnamefont {Zhang}}, \bibinfo {author}
  {\bibfnamefont {I.~R.}\ \bibnamefont {Fisher}}, \bibinfo {author}
  {\bibfnamefont {Z.}~\bibnamefont {Hussain}}, \ and\ \bibinfo {author}
  {\bibfnamefont {Z.-X.}\ \bibnamefont {Shen}},\ }\href {\doibase
  10.1126/science.1173034} {\bibfield  {journal} {\bibinfo  {journal}
  {Science}\ }\textbf {\bibinfo {volume} {325}},\ \bibinfo {pages} {178}
  (\bibinfo {year} {2009})}\BibitemShut {NoStop}%
\bibitem [{\citenamefont {Bernevig}\ \emph {et~al.}(2006)\citenamefont
  {Bernevig}, \citenamefont {Hughes},\ and\ \citenamefont
  {Zhang}}]{Bernevig2006}%
  \BibitemOpen
  \bibfield  {author} {\bibinfo {author} {\bibfnamefont {B.~A.}\ \bibnamefont
  {Bernevig}}, \bibinfo {author} {\bibfnamefont {T.~L.}\ \bibnamefont
  {Hughes}}, \ and\ \bibinfo {author} {\bibfnamefont {S.-C.}\ \bibnamefont
  {Zhang}},\ }\href {\doibase 10.1126/science.1133734} {\bibfield  {journal}
  {\bibinfo  {journal} {Science}\ }\textbf {\bibinfo {volume} {314}},\ \bibinfo
  {pages} {1757} (\bibinfo {year} {2006})}\BibitemShut {NoStop}%
\bibitem [{\citenamefont {K{\"o}nig}\ \emph {et~al.}(2007)\citenamefont
  {K{\"o}nig}, \citenamefont {Wiedmann}, \citenamefont {Br{\"u}ne},
  \citenamefont {Roth}, \citenamefont {Buhmann}, \citenamefont {Molenkamp},
  \citenamefont {Qi},\ and\ \citenamefont {Zhang}}]{Konig2007}%
  \BibitemOpen
  \bibfield  {author} {\bibinfo {author} {\bibfnamefont {M.}~\bibnamefont
  {K{\"o}nig}}, \bibinfo {author} {\bibfnamefont {S.}~\bibnamefont {Wiedmann}},
  \bibinfo {author} {\bibfnamefont {C.}~\bibnamefont {Br{\"u}ne}}, \bibinfo
  {author} {\bibfnamefont {A.}~\bibnamefont {Roth}}, \bibinfo {author}
  {\bibfnamefont {H.}~\bibnamefont {Buhmann}}, \bibinfo {author} {\bibfnamefont
  {L.~W.}\ \bibnamefont {Molenkamp}}, \bibinfo {author} {\bibfnamefont {X.-L.}\
  \bibnamefont {Qi}}, \ and\ \bibinfo {author} {\bibfnamefont {S.-C.}\
  \bibnamefont {Zhang}},\ }\href {\doibase 10.1126/science.1148047} {\bibfield
  {journal} {\bibinfo  {journal} {Science}\ }\textbf {\bibinfo {volume}
  {318}},\ \bibinfo {pages} {766} (\bibinfo {year} {2007})}\BibitemShut
  {NoStop}%
\bibitem [{\citenamefont {Haldane}\ and\ \citenamefont
  {Raghu}(2008)}]{Haldane2008}%
  \BibitemOpen
  \bibfield  {author} {\bibinfo {author} {\bibfnamefont {F.~D.~M.}\
  \bibnamefont {Haldane}}\ and\ \bibinfo {author} {\bibfnamefont
  {S.}~\bibnamefont {Raghu}},\ }\href {\doibase 10.1103/PhysRevLett.100.013904}
  {\bibfield  {journal} {\bibinfo  {journal} {Phys. Rev. Lett.}\ }\textbf
  {\bibinfo {volume} {100}},\ \bibinfo {pages} {013904} (\bibinfo {year}
  {2008})}\BibitemShut {NoStop}%
\bibitem [{\citenamefont {Wang}\ \emph {et~al.}(2009)\citenamefont {Wang},
  \citenamefont {Chong}, \citenamefont {Joannopoulos},\ and\ \citenamefont
  {Soljacic}}]{Wang_Nature2009}%
  \BibitemOpen
  \bibfield  {author} {\bibinfo {author} {\bibfnamefont {Z.}~\bibnamefont
  {Wang}}, \bibinfo {author} {\bibfnamefont {Y.}~\bibnamefont {Chong}},
  \bibinfo {author} {\bibfnamefont {J.~D.}\ \bibnamefont {Joannopoulos}}, \
  and\ \bibinfo {author} {\bibfnamefont {M.}~\bibnamefont {Soljacic}},\ }\href
  {\doibase 10.1038/nature08293} {\bibfield  {journal} {\bibinfo  {journal}
  {Nature}\ }\textbf {\bibinfo {volume} {461}},\ \bibinfo {pages} {772}
  (\bibinfo {year} {2009})}\BibitemShut {NoStop}%
\bibitem [{\citenamefont {Khanikaev}\ \emph {et~al.}(2013)\citenamefont
  {Khanikaev}, \citenamefont {Hossein~Mousavi}, \citenamefont {Tse},
  \citenamefont {Kargarian}, \citenamefont {MacDonald},\ and\ \citenamefont
  {Shvets}}]{Khanikaev2013}%
  \BibitemOpen
  \bibfield  {author} {\bibinfo {author} {\bibfnamefont {A.~B.}\ \bibnamefont
  {Khanikaev}}, \bibinfo {author} {\bibfnamefont {S.}~\bibnamefont
  {Hossein~Mousavi}}, \bibinfo {author} {\bibfnamefont {W.-K.}\ \bibnamefont
  {Tse}}, \bibinfo {author} {\bibfnamefont {M.}~\bibnamefont {Kargarian}},
  \bibinfo {author} {\bibfnamefont {A.~H.}\ \bibnamefont {MacDonald}}, \ and\
  \bibinfo {author} {\bibfnamefont {G.}~\bibnamefont {Shvets}},\ }\href
  {\doibase 10.1038/nmat3520} {\bibfield  {journal} {\bibinfo  {journal}
  {Nature Materials}\ }\textbf {\bibinfo {volume} {12}},\ \bibinfo {pages}
  {233} (\bibinfo {year} {2013})}\BibitemShut {NoStop}%
\bibitem [{\citenamefont {Karzig}\ \emph {et~al.}(2015)\citenamefont {Karzig},
  \citenamefont {Bardyn}, \citenamefont {Lindner},\ and\ \citenamefont
  {Refael}}]{Karzig-PRX-2015}%
  \BibitemOpen
  \bibfield  {author} {\bibinfo {author} {\bibfnamefont {T.}~\bibnamefont
  {Karzig}}, \bibinfo {author} {\bibfnamefont {C.-E.}\ \bibnamefont {Bardyn}},
  \bibinfo {author} {\bibfnamefont {N.~H.}\ \bibnamefont {Lindner}}, \ and\
  \bibinfo {author} {\bibfnamefont {G.}~\bibnamefont {Refael}},\ }\href
  {\doibase 10.1103/PhysRevX.5.031001} {\bibfield  {journal} {\bibinfo
  {journal} {Phys. Rev. X}\ }\textbf {\bibinfo {volume} {5}},\ \bibinfo {pages}
  {031001} (\bibinfo {year} {2015})}\BibitemShut {NoStop}%
\bibitem [{\citenamefont {Bardyn}\ \emph {et~al.}(2015)\citenamefont {Bardyn},
  \citenamefont {Karzig}, \citenamefont {Refael},\ and\ \citenamefont
  {Liew}}]{Bardyn-PRB-2015}%
  \BibitemOpen
  \bibfield  {author} {\bibinfo {author} {\bibfnamefont {C.-E.}\ \bibnamefont
  {Bardyn}}, \bibinfo {author} {\bibfnamefont {T.}~\bibnamefont {Karzig}},
  \bibinfo {author} {\bibfnamefont {G.}~\bibnamefont {Refael}}, \ and\ \bibinfo
  {author} {\bibfnamefont {T.~C.~H.}\ \bibnamefont {Liew}},\ }\href {\doibase
  10.1103/PhysRevB.91.161413} {\bibfield  {journal} {\bibinfo  {journal} {Phys.
  Rev. B}\ }\textbf {\bibinfo {volume} {91}},\ \bibinfo {pages} {161413(R)}
  (\bibinfo {year} {2015})}\BibitemShut {NoStop}%
\bibitem [{\citenamefont {Nalitov}\ \emph {et~al.}(2015)\citenamefont
  {Nalitov}, \citenamefont {Solnyshkov},\ and\ \citenamefont
  {Malpuech}}]{Nalitov-Z}%
  \BibitemOpen
  \bibfield  {author} {\bibinfo {author} {\bibfnamefont {A.~V.}\ \bibnamefont
  {Nalitov}}, \bibinfo {author} {\bibfnamefont {D.~D.}\ \bibnamefont
  {Solnyshkov}}, \ and\ \bibinfo {author} {\bibfnamefont {G.}~\bibnamefont
  {Malpuech}},\ }\href {\doibase 10.1103/PhysRevLett.114.116401} {\bibfield
  {journal} {\bibinfo  {journal} {Phys. Rev. Lett.}\ }\textbf {\bibinfo
  {volume} {114}},\ \bibinfo {pages} {116401} (\bibinfo {year}
  {2015})}\BibitemShut {NoStop}%
\bibitem [{\citenamefont {Gulevich}\ \emph {et~al.}(2016)\citenamefont
  {Gulevich}, \citenamefont {Yudin}, \citenamefont {Iorsh},\ and\ \citenamefont
  {Shelykh}}]{KagomePolariton}%
  \BibitemOpen
  \bibfield  {author} {\bibinfo {author} {\bibfnamefont {D.~R.}\ \bibnamefont
  {Gulevich}}, \bibinfo {author} {\bibfnamefont {D.}~\bibnamefont {Yudin}},
  \bibinfo {author} {\bibfnamefont {I.~V.}\ \bibnamefont {Iorsh}}, \ and\
  \bibinfo {author} {\bibfnamefont {I.~A.}\ \bibnamefont {Shelykh}},\ }\href
  {\doibase 10.1103/PhysRevB.94.115437} {\bibfield  {journal} {\bibinfo
  {journal} {Phys. Rev. B}\ }\textbf {\bibinfo {volume} {94}},\ \bibinfo
  {pages} {115437} (\bibinfo {year} {2016})}\BibitemShut {NoStop}%
\bibitem [{\citenamefont {Kozin}\ \emph {et~al.}(2018)\citenamefont {Kozin},
  \citenamefont {Shelykh}, \citenamefont {Nalitov},\ and\ \citenamefont
  {Iorsh}}]{Kozin2018}%
  \BibitemOpen
  \bibfield  {author} {\bibinfo {author} {\bibfnamefont {V.~K.}\ \bibnamefont
  {Kozin}}, \bibinfo {author} {\bibfnamefont {I.~A.}\ \bibnamefont {Shelykh}},
  \bibinfo {author} {\bibfnamefont {A.~V.}\ \bibnamefont {Nalitov}}, \ and\
  \bibinfo {author} {\bibfnamefont {I.~V.}\ \bibnamefont {Iorsh}},\ }\href
  {\doibase 10.1103/PhysRevB.98.125115} {\bibfield  {journal} {\bibinfo
  {journal} {Phys. Rev. B}\ }\textbf {\bibinfo {volume} {98}},\ \bibinfo
  {pages} {125115} (\bibinfo {year} {2018})}\BibitemShut {NoStop}%
\bibitem [{\citenamefont {Klembt}\ \emph {et~al.}(2018)\citenamefont {Klembt},
  \citenamefont {Harder}, \citenamefont {Egorov}, \citenamefont {Winkler},
  \citenamefont {Ge}, \citenamefont {Bandres}, \citenamefont {Emmerling},
  \citenamefont {Worschech}, \citenamefont {Liew}, \citenamefont {Segev},
  \citenamefont {Schneider},\ and\ \citenamefont
  {H{\"o}fling}}]{Klembt_Nature2018}%
  \BibitemOpen
  \bibfield  {author} {\bibinfo {author} {\bibfnamefont {S.}~\bibnamefont
  {Klembt}}, \bibinfo {author} {\bibfnamefont {T.~H.}\ \bibnamefont {Harder}},
  \bibinfo {author} {\bibfnamefont {O.~A.}\ \bibnamefont {Egorov}}, \bibinfo
  {author} {\bibfnamefont {K.}~\bibnamefont {Winkler}}, \bibinfo {author}
  {\bibfnamefont {R.}~\bibnamefont {Ge}}, \bibinfo {author} {\bibfnamefont
  {M.~A.}\ \bibnamefont {Bandres}}, \bibinfo {author} {\bibfnamefont
  {M.}~\bibnamefont {Emmerling}}, \bibinfo {author} {\bibfnamefont
  {L.}~\bibnamefont {Worschech}}, \bibinfo {author} {\bibfnamefont {T.~C.~H.}\
  \bibnamefont {Liew}}, \bibinfo {author} {\bibfnamefont {M.}~\bibnamefont
  {Segev}}, \bibinfo {author} {\bibfnamefont {C.}~\bibnamefont {Schneider}}, \
  and\ \bibinfo {author} {\bibfnamefont {S.}~\bibnamefont {H{\"o}fling}},\
  }\href {\doibase 10.1038/s41586-018-0601-5} {\bibfield  {journal} {\bibinfo
  {journal} {Nature}\ }\textbf {\bibinfo {volume} {562}},\ \bibinfo {pages}
  {552} (\bibinfo {year} {2018})}\BibitemShut {NoStop}%
\bibitem [{\citenamefont {Bleu}\ \emph {et~al.}(2016)\citenamefont {Bleu},
  \citenamefont {Solnyshkov},\ and\ \citenamefont {Malpuech}}]{Bleu2016}%
  \BibitemOpen
  \bibfield  {author} {\bibinfo {author} {\bibfnamefont {O.}~\bibnamefont
  {Bleu}}, \bibinfo {author} {\bibfnamefont {D.~D.}\ \bibnamefont
  {Solnyshkov}}, \ and\ \bibinfo {author} {\bibfnamefont {G.}~\bibnamefont
  {Malpuech}},\ }\href {\doibase 10.1103/PhysRevB.93.085438} {\bibfield
  {journal} {\bibinfo  {journal} {Phys. Rev. B}\ }\textbf {\bibinfo {volume}
  {93}},\ \bibinfo {pages} {85438} (\bibinfo {year} {2016})}\BibitemShut
  {NoStop}%
\bibitem [{\citenamefont {Bardyn}\ \emph {et~al.}(2016)\citenamefont {Bardyn},
  \citenamefont {Karzig}, \citenamefont {Refael},\ and\ \citenamefont
  {Liew}}]{Bardyn_2016PRB}%
  \BibitemOpen
  \bibfield  {author} {\bibinfo {author} {\bibfnamefont {C.-E.}\ \bibnamefont
  {Bardyn}}, \bibinfo {author} {\bibfnamefont {T.}~\bibnamefont {Karzig}},
  \bibinfo {author} {\bibfnamefont {G.}~\bibnamefont {Refael}}, \ and\ \bibinfo
  {author} {\bibfnamefont {T.~C.~H.}\ \bibnamefont {Liew}},\ }\href {\doibase
  10.1103/PhysRevB.93.020502} {\bibfield  {journal} {\bibinfo  {journal} {Phys.
  Rev. B}\ }\textbf {\bibinfo {volume} {93}},\ \bibinfo {pages} {020502(R)}
  (\bibinfo {year} {2016})}\BibitemShut {NoStop}%
\bibitem [{\citenamefont {Gulevich}\ \emph {et~al.}(2017)\citenamefont
  {Gulevich}, \citenamefont {Yudin}, \citenamefont {Skryabin}, \citenamefont
  {Iorsh},\ and\ \citenamefont {Shelykh}}]{Gulevich2017}%
  \BibitemOpen
  \bibfield  {author} {\bibinfo {author} {\bibfnamefont {D.~R.}\ \bibnamefont
  {Gulevich}}, \bibinfo {author} {\bibfnamefont {D.}~\bibnamefont {Yudin}},
  \bibinfo {author} {\bibfnamefont {D.~V.}\ \bibnamefont {Skryabin}}, \bibinfo
  {author} {\bibfnamefont {I.~V.}\ \bibnamefont {Iorsh}}, \ and\ \bibinfo
  {author} {\bibfnamefont {I.~A.}\ \bibnamefont {Shelykh}},\ }\href {\doibase
  10.1038/s41598-017-01646-y} {\bibfield  {journal} {\bibinfo  {journal}
  {Scientific Reports}\ }\textbf {\bibinfo {volume} {7}},\ \bibinfo {pages}
  {1780} (\bibinfo {year} {2017})}\BibitemShut {NoStop}%
\bibitem [{\citenamefont {Sigurdsson}\ \emph
  {et~al.}(2017{\natexlab{a}})\citenamefont {Sigurdsson}, \citenamefont {Li},\
  and\ \citenamefont {Liew}}]{Sigurdsson2017}%
  \BibitemOpen
  \bibfield  {author} {\bibinfo {author} {\bibfnamefont {H.}~\bibnamefont
  {Sigurdsson}}, \bibinfo {author} {\bibfnamefont {G.}~\bibnamefont {Li}}, \
  and\ \bibinfo {author} {\bibfnamefont {T.~C.~H.}\ \bibnamefont {Liew}},\
  }\href {\doibase 10.1103/PhysRevB.96.115453} {\bibfield  {journal} {\bibinfo
  {journal} {Phys. Rev. B}\ }\textbf {\bibinfo {volume} {96}},\ \bibinfo
  {pages} {115453} (\bibinfo {year} {2017}{\natexlab{a}})}\BibitemShut
  {NoStop}%
\bibitem [{\citenamefont {Kartashov}\ and\ \citenamefont
  {Skryabin}(2017)}]{Kartashov2017}%
  \BibitemOpen
  \bibfield  {author} {\bibinfo {author} {\bibfnamefont {Y.~V.}\ \bibnamefont
  {Kartashov}}\ and\ \bibinfo {author} {\bibfnamefont {D.~V.}\ \bibnamefont
  {Skryabin}},\ }\href {\doibase 10.1103/PhysRevLett.119.253904} {\bibfield
  {journal} {\bibinfo  {journal} {Phys. Rev. Lett.}\ }\textbf {\bibinfo
  {volume} {119}},\ \bibinfo {pages} {253904} (\bibinfo {year}
  {2017})}\BibitemShut {NoStop}%
\bibitem [{\citenamefont {Bleu}\ \emph {et~al.}(2017)\citenamefont {Bleu},
  \citenamefont {Solnyshkov},\ and\ \citenamefont {Malpuech}}]{Bleu_PRB2017}%
  \BibitemOpen
  \bibfield  {author} {\bibinfo {author} {\bibfnamefont {O.}~\bibnamefont
  {Bleu}}, \bibinfo {author} {\bibfnamefont {D.~D.}\ \bibnamefont
  {Solnyshkov}}, \ and\ \bibinfo {author} {\bibfnamefont {G.}~\bibnamefont
  {Malpuech}},\ }\href {\doibase 10.1103/PhysRevB.95.115415} {\bibfield
  {journal} {\bibinfo  {journal} {Phys. Rev. B}\ }\textbf {\bibinfo {volume}
  {95}},\ \bibinfo {pages} {115415} (\bibinfo {year} {2017})}\BibitemShut
  {NoStop}%
\bibitem [{\citenamefont {Mandal}\ \emph {et~al.}(2019)\citenamefont {Mandal},
  \citenamefont {Ge},\ and\ \citenamefont {Liew}}]{Mandal_PRB2019}%
  \BibitemOpen
  \bibfield  {author} {\bibinfo {author} {\bibfnamefont {S.}~\bibnamefont
  {Mandal}}, \bibinfo {author} {\bibfnamefont {R.}~\bibnamefont {Ge}}, \ and\
  \bibinfo {author} {\bibfnamefont {T.~C.~H.}\ \bibnamefont {Liew}},\ }\href
  {\doibase 10.1103/PhysRevB.99.115423} {\bibfield  {journal} {\bibinfo
  {journal} {Phys. Rev. B}\ }\textbf {\bibinfo {volume} {99}},\ \bibinfo
  {pages} {115423} (\bibinfo {year} {2019})}\BibitemShut {NoStop}%
\bibitem [{\citenamefont {St-Jean}\ \emph {et~al.}(2017)\citenamefont
  {St-Jean}, \citenamefont {Goblot}, \citenamefont {Galopin}, \citenamefont
  {Lema{\^i}tre}, \citenamefont {Ozawa}, \citenamefont {Le~Gratiet},
  \citenamefont {Sagnes}, \citenamefont {Bloch},\ and\ \citenamefont
  {Amo}}]{StJean_NatPho2017}%
  \BibitemOpen
  \bibfield  {author} {\bibinfo {author} {\bibfnamefont {P.}~\bibnamefont
  {St-Jean}}, \bibinfo {author} {\bibfnamefont {V.}~\bibnamefont {Goblot}},
  \bibinfo {author} {\bibfnamefont {E.}~\bibnamefont {Galopin}}, \bibinfo
  {author} {\bibfnamefont {A.}~\bibnamefont {Lema{\^i}tre}}, \bibinfo {author}
  {\bibfnamefont {T.}~\bibnamefont {Ozawa}}, \bibinfo {author} {\bibfnamefont
  {L.}~\bibnamefont {Le~Gratiet}}, \bibinfo {author} {\bibfnamefont
  {I.}~\bibnamefont {Sagnes}}, \bibinfo {author} {\bibfnamefont
  {J.}~\bibnamefont {Bloch}}, \ and\ \bibinfo {author} {\bibfnamefont
  {A.}~\bibnamefont {Amo}},\ }\href {\doibase 10.1038/s41566-017-0006-2}
  {\bibfield  {journal} {\bibinfo  {journal} {Nature Photonics}\ }\textbf
  {\bibinfo {volume} {11}},\ \bibinfo {pages} {651} (\bibinfo {year}
  {2017})}\BibitemShut {NoStop}%
\bibitem [{\citenamefont {Kartashov}\ and\ \citenamefont
  {Skryabin}(2019)}]{Kartashov_PRL2019}%
  \BibitemOpen
  \bibfield  {author} {\bibinfo {author} {\bibfnamefont {Y.~V.}\ \bibnamefont
  {Kartashov}}\ and\ \bibinfo {author} {\bibfnamefont {D.~V.}\ \bibnamefont
  {Skryabin}},\ }\href {\doibase 10.1103/PhysRevLett.122.083902} {\bibfield
  {journal} {\bibinfo  {journal} {Phys. Rev. Lett.}\ }\textbf {\bibinfo
  {volume} {122}},\ \bibinfo {pages} {083902} (\bibinfo {year}
  {2019})}\BibitemShut {NoStop}%
\bibitem [{\citenamefont {Brunetti}\ \emph {et~al.}(2006)\citenamefont
  {Brunetti}, \citenamefont {Vladimirova}, \citenamefont {Scalbert},
  \citenamefont {Andr\'e}, \citenamefont {Solnyshkov}, \citenamefont
  {Malpuech}, \citenamefont {Shelykh},\ and\ \citenamefont
  {Kavokin}}]{Brunetti2006}%
  \BibitemOpen
  \bibfield  {author} {\bibinfo {author} {\bibfnamefont {A.}~\bibnamefont
  {Brunetti}}, \bibinfo {author} {\bibfnamefont {M.}~\bibnamefont
  {Vladimirova}}, \bibinfo {author} {\bibfnamefont {D.}~\bibnamefont
  {Scalbert}}, \bibinfo {author} {\bibfnamefont {R.}~\bibnamefont {Andr\'e}},
  \bibinfo {author} {\bibfnamefont {D.}~\bibnamefont {Solnyshkov}}, \bibinfo
  {author} {\bibfnamefont {G.}~\bibnamefont {Malpuech}}, \bibinfo {author}
  {\bibfnamefont {I.~A.}\ \bibnamefont {Shelykh}}, \ and\ \bibinfo {author}
  {\bibfnamefont {A.~V.}\ \bibnamefont {Kavokin}},\ }\href {\doibase
  10.1103/PhysRevB.73.205337} {\bibfield  {journal} {\bibinfo  {journal} {Phys.
  Rev. B}\ }\textbf {\bibinfo {volume} {73}},\ \bibinfo {pages} {205337}
  (\bibinfo {year} {2006})}\BibitemShut {NoStop}%
\bibitem [{\citenamefont {Kr{\'o}l}\ \emph {et~al.}(2018)\citenamefont
  {Kr{\'o}l}, \citenamefont {Mirek}, \citenamefont {Lekenta}, \citenamefont
  {Rousset}, \citenamefont {Stephan}, \citenamefont {Nawrocki}, \citenamefont
  {Matuszewski}, \citenamefont {Szczytko}, \citenamefont {Pacuski},\ and\
  \citenamefont {Pietka}}]{Krol2018}%
  \BibitemOpen
  \bibfield  {author} {\bibinfo {author} {\bibfnamefont {M.}~\bibnamefont
  {Kr{\'o}l}}, \bibinfo {author} {\bibfnamefont {R.}~\bibnamefont {Mirek}},
  \bibinfo {author} {\bibfnamefont {K.}~\bibnamefont {Lekenta}}, \bibinfo
  {author} {\bibfnamefont {J.-G.}\ \bibnamefont {Rousset}}, \bibinfo {author}
  {\bibfnamefont {D.}~\bibnamefont {Stephan}}, \bibinfo {author} {\bibfnamefont
  {M.}~\bibnamefont {Nawrocki}}, \bibinfo {author} {\bibfnamefont
  {M.}~\bibnamefont {Matuszewski}}, \bibinfo {author} {\bibfnamefont
  {J.}~\bibnamefont {Szczytko}}, \bibinfo {author} {\bibfnamefont
  {W.}~\bibnamefont {Pacuski}}, \ and\ \bibinfo {author} {\bibfnamefont
  {B.}~\bibnamefont {Pietka}},\ }\href {\doibase 10.1038/s41598-018-25018-2}
  {\bibfield  {journal} {\bibinfo  {journal} {Scientific Reports}\ }\textbf
  {\bibinfo {volume} {8}},\ \bibinfo {pages} {6694} (\bibinfo {year}
  {2018})}\BibitemShut {NoStop}%
\bibitem [{\citenamefont {Ohadi}\ \emph {et~al.}(2015)\citenamefont {Ohadi},
  \citenamefont {Dreismann}, \citenamefont {Rubo}, \citenamefont {Pinsker},
  \citenamefont {del Valle-Inclan~Redondo}, \citenamefont {Tsintzos},
  \citenamefont {Hatzopoulos}, \citenamefont {Savvidis},\ and\ \citenamefont
  {Baumberg}}]{Ohadi2015}%
  \BibitemOpen
  \bibfield  {author} {\bibinfo {author} {\bibfnamefont {H.}~\bibnamefont
  {Ohadi}}, \bibinfo {author} {\bibfnamefont {A.}~\bibnamefont {Dreismann}},
  \bibinfo {author} {\bibfnamefont {Y.~G.}\ \bibnamefont {Rubo}}, \bibinfo
  {author} {\bibfnamefont {F.}~\bibnamefont {Pinsker}}, \bibinfo {author}
  {\bibfnamefont {Y.}~\bibnamefont {del Valle-Inclan~Redondo}}, \bibinfo
  {author} {\bibfnamefont {S.~I.}\ \bibnamefont {Tsintzos}}, \bibinfo {author}
  {\bibfnamefont {Z.}~\bibnamefont {Hatzopoulos}}, \bibinfo {author}
  {\bibfnamefont {P.~G.}\ \bibnamefont {Savvidis}}, \ and\ \bibinfo {author}
  {\bibfnamefont {J.~J.}\ \bibnamefont {Baumberg}},\ }\href {\doibase
  10.1103/PhysRevX.5.031002} {\bibfield  {journal} {\bibinfo  {journal} {Phys.
  Rev. X}\ }\textbf {\bibinfo {volume} {5}},\ \bibinfo {pages} {031002}
  (\bibinfo {year} {2015})}\BibitemShut {NoStop}%
\bibitem [{\citenamefont {Ohadi}\ \emph {et~al.}(2016)\citenamefont {Ohadi},
  \citenamefont {Gregory}, \citenamefont {Freegarde}, \citenamefont {Rubo},
  \citenamefont {Kavokin}, \citenamefont {Berloff},\ and\ \citenamefont
  {Lagoudakis}}]{Ohadi2016}%
  \BibitemOpen
  \bibfield  {author} {\bibinfo {author} {\bibfnamefont {H.}~\bibnamefont
  {Ohadi}}, \bibinfo {author} {\bibfnamefont {R.~L.}\ \bibnamefont {Gregory}},
  \bibinfo {author} {\bibfnamefont {T.}~\bibnamefont {Freegarde}}, \bibinfo
  {author} {\bibfnamefont {Y.~G.}\ \bibnamefont {Rubo}}, \bibinfo {author}
  {\bibfnamefont {A.~V.}\ \bibnamefont {Kavokin}}, \bibinfo {author}
  {\bibfnamefont {N.~G.}\ \bibnamefont {Berloff}}, \ and\ \bibinfo {author}
  {\bibfnamefont {P.~G.}\ \bibnamefont {Lagoudakis}},\ }\href {\doibase
  10.1103/PhysRevX.6.031032} {\bibfield  {journal} {\bibinfo  {journal}
  {Physical Review X}\ }\textbf {\bibinfo {volume} {6}},\ \bibinfo {pages}
  {031032} (\bibinfo {year} {2016})}\BibitemShut {NoStop}%
\bibitem [{\citenamefont {Dreismann}\ \emph {et~al.}(2016)\citenamefont
  {Dreismann}, \citenamefont {Ohadi}, \citenamefont {del Valle-Inclan~Redondo},
  \citenamefont {Balili}, \citenamefont {Rubo}, \citenamefont {Tsintzos},
  \citenamefont {Deligeorgis}, \citenamefont {Hatzopoulos}, \citenamefont
  {Savvidis},\ and\ \citenamefont {Baumberg}}]{Dreismann2016}%
  \BibitemOpen
  \bibfield  {author} {\bibinfo {author} {\bibfnamefont {A.}~\bibnamefont
  {Dreismann}}, \bibinfo {author} {\bibfnamefont {H.}~\bibnamefont {Ohadi}},
  \bibinfo {author} {\bibfnamefont {Y.}~\bibnamefont {del
  Valle-Inclan~Redondo}}, \bibinfo {author} {\bibfnamefont {R.}~\bibnamefont
  {Balili}}, \bibinfo {author} {\bibfnamefont {Y.~G.}\ \bibnamefont {Rubo}},
  \bibinfo {author} {\bibfnamefont {S.~I.}\ \bibnamefont {Tsintzos}}, \bibinfo
  {author} {\bibfnamefont {G.}~\bibnamefont {Deligeorgis}}, \bibinfo {author}
  {\bibfnamefont {Z.}~\bibnamefont {Hatzopoulos}}, \bibinfo {author}
  {\bibfnamefont {P.~G.}\ \bibnamefont {Savvidis}}, \ and\ \bibinfo {author}
  {\bibfnamefont {J.~J.}\ \bibnamefont {Baumberg}},\ }\href {\doibase
  10.1038/nmat4722} {\bibfield  {journal} {\bibinfo  {journal} {Nature
  Materials}\ }\textbf {\bibinfo {volume} {15}},\ \bibinfo {pages} {1074}
  (\bibinfo {year} {2016})}\BibitemShut {NoStop}%
\bibitem [{\citenamefont {Ohadi}\ \emph {et~al.}(2017)\citenamefont {Ohadi},
  \citenamefont {Ramsay}, \citenamefont {Sigurdsson}, \citenamefont {del
  Valle-Inclan~Redondo}, \citenamefont {Tsintzos}, \citenamefont {Hatzopoulos},
  \citenamefont {Liew}, \citenamefont {Shelykh}, \citenamefont {Rubo},
  \citenamefont {Savvidis},\ and\ \citenamefont {Baumberg}}]{Ohadi2017}%
  \BibitemOpen
  \bibfield  {author} {\bibinfo {author} {\bibfnamefont {H.}~\bibnamefont
  {Ohadi}}, \bibinfo {author} {\bibfnamefont {A.~J.}\ \bibnamefont {Ramsay}},
  \bibinfo {author} {\bibfnamefont {H.}~\bibnamefont {Sigurdsson}}, \bibinfo
  {author} {\bibfnamefont {Y.}~\bibnamefont {del Valle-Inclan~Redondo}},
  \bibinfo {author} {\bibfnamefont {S.~I.}\ \bibnamefont {Tsintzos}}, \bibinfo
  {author} {\bibfnamefont {Z.}~\bibnamefont {Hatzopoulos}}, \bibinfo {author}
  {\bibfnamefont {T.~C.~H.}\ \bibnamefont {Liew}}, \bibinfo {author}
  {\bibfnamefont {I.~A.}\ \bibnamefont {Shelykh}}, \bibinfo {author}
  {\bibfnamefont {Y.~G.}\ \bibnamefont {Rubo}}, \bibinfo {author}
  {\bibfnamefont {P.~G.}\ \bibnamefont {Savvidis}}, \ and\ \bibinfo {author}
  {\bibfnamefont {J.~J.}\ \bibnamefont {Baumberg}},\ }\href {\doibase
  10.1103/PhysRevLett.119.067401} {\bibfield  {journal} {\bibinfo  {journal}
  {Phys. Rev. Lett.}\ }\textbf {\bibinfo {volume} {119}},\ \bibinfo {pages}
  {067401} (\bibinfo {year} {2017})}\BibitemShut {NoStop}%
\bibitem [{\citenamefont {Sigurdsson}\ \emph
  {et~al.}(2017{\natexlab{b}})\citenamefont {Sigurdsson}, \citenamefont
  {Ramsay}, \citenamefont {Ohadi}, \citenamefont {Rubo}, \citenamefont {Liew},
  \citenamefont {Baumberg},\ and\ \citenamefont {Shelykh}}]{Sigurdsson2017a}%
  \BibitemOpen
  \bibfield  {author} {\bibinfo {author} {\bibfnamefont {H.}~\bibnamefont
  {Sigurdsson}}, \bibinfo {author} {\bibfnamefont {A.~J.}\ \bibnamefont
  {Ramsay}}, \bibinfo {author} {\bibfnamefont {H.}~\bibnamefont {Ohadi}},
  \bibinfo {author} {\bibfnamefont {Y.~G.}\ \bibnamefont {Rubo}}, \bibinfo
  {author} {\bibfnamefont {T.~C.~H.}\ \bibnamefont {Liew}}, \bibinfo {author}
  {\bibfnamefont {J.~J.}\ \bibnamefont {Baumberg}}, \ and\ \bibinfo {author}
  {\bibfnamefont {I.~A.}\ \bibnamefont {Shelykh}},\ }\href {\doibase
  10.1103/PhysRevB.96.155403} {\bibfield  {journal} {\bibinfo  {journal} {Phys.
  Rev. B}\ }\textbf {\bibinfo {volume} {96}},\ \bibinfo {pages} {155403}
  (\bibinfo {year} {2017}{\natexlab{b}})}\BibitemShut {NoStop}%
\bibitem [{Sup()}]{Supplementary2019}%
  \BibitemOpen
  \href@noop {} {}\bibinfo {note} {See Supplemental Material}\BibitemShut
  {NoStop}%
\end{thebibliography}
\end{document}